\documentclass[aps,pra,reprint,floatfix]{revtex4-1}

\usepackage{color}
\usepackage{amsmath}
\usepackage{mathtools}
\usepackage{graphicx}
\usepackage{braket}
\usepackage{dcolumn}
\usepackage{bm}
\usepackage{comment}
\usepackage{floatrow}
\usepackage[caption=false]{subfig}
\usepackage{bbold}
\usepackage{bbm}
\usepackage{esint}
\usepackage{overpic}
\usepackage[colorlinks=true,citecolor=blue,linkcolor=magenta]{hyperref}
\usepackage{cleveref}

\newcommand{\rmd}{\mathrm{d}}
\newcommand{\rmi}{\mathrm{i}}
\newcommand{\rme}{\mathrm{e}}

\crefname{equation}{Eq.}{Eqs.}
\Crefname{equation}{Equation}{Equations}
\crefname{figure}{Fig.}{Figs.}
\Crefname{figure}{Figure}{Figures}

\begin{document}
 
\title{Strongly correlated photon transport in nonlinear photonic lattices with disorder: Probing signatures of the localisation transition}

\author{Tian Feng See\textsuperscript{1}}
\email{baise.feng@gmail.com}
\author{V. M. Bastidas\textsuperscript{2}}
\author{Jirawat Tangpanitanon\textsuperscript{1}}
\author{Dimitris G. Angelakis\textsuperscript{1,3}}
\email{dimitris.angelakis@gmail.com}
\affiliation{\textsuperscript{1}%
	Centre for Quantum Technologies, National University of Singapore, 3 Science Drive 2, Singapore 117543, Singapore%
}
  \affiliation{\textsuperscript{2}%
  NTT Basic Research Laboratories \& Research Center for Theoretical Quantum Physics, 3-1 Morinosato-Wakamiya, Atsugi, Kanagawa, 243-0198, Japan
  } 
\affiliation{\textsuperscript{3}%
	School of Electrical and Computer Engineering, Technical University of Crete, Chania, Crete, 73100 Greece%
}
\date{\today}

\begin{abstract}
We study the transport of few-photon states in open disordered nonlinear photonic lattices. More specifically, we consider a waveguide quantum electrodynamics (QED) setup where photons are scattered from a chain of nonlinear resonators with on-site Bose-Hubbard interaction in the presence of an incommensurate potential. Applying our recently developed diagrammatic technique that evaluates the scattering matrix (S-matrix) via absorption and emission diagrams, we compute the two-photon transmission probability and show that it carries signatures of the underlying localisation transition of the system. We compare the calculated probability to the participation ratio of the eigenstates and find close agreement for a range of interaction strengths. The scaling of the two-photon transmission probability suggests that there might be two localisation transitions in the high energy eigenstates corresponding to interaction and quasiperiodicity respectively. This observation is absent from the participation ratio. We analyse the robustness of the transmission signatures against local dissipation and briefly discuss possible implementation using current technology.
\end{abstract}

\maketitle

\section{Introduction}\label{sec:introduction}
Recent advances in quantum nonlinear optics and circuit QED have allowed the engineering of few-body photonic states exhibiting signatures of many-body effects \cite{noh16,hartmann16,angelakis17,roy17}. Experimentally demonstrated effects using superconducting circuits include dissipative phase transition in a chain of nonlinear QED resonators as well as many-body localisation transition \cite{fitzpatrick17,roushan17}. Strongly correlated states of light have also been created in slow light Rydberg polaritons \cite{firstenberg13}, and excitonic systems are progressing towards this direction as well \cite{amo16}. An important aspect of many-body photonic simulators is that they are open optical systems. Performing spectroscopy will thus require an analysis of the photon transmission spectra and statistics as was done for the Bose-Hubbard model in recent works \cite{lee15,see17,pedersen17}. In quantum optics, this generally assumes scattering photons from the system via waveguides. Waveguide QED setups for quantum technology applications have been widely proposed, and experimentally implemented, usually containing a multilevel emitter or few uncorrelated two-level emitters  \cite{lodahl17,chang14,reiserer15,lodahl15,mitsch14,goban15,sollner15,young15,scheucher16,coles16,gu17}.

To treat the problem of few-photon scattering from quantum emitters, one needs to use a combination of the Lippmann-Schwinger formalism, or equivalently, the Bethe ansatz approach, and the input-output formalism from quantum optics \cite{shen07,zheng10,roy11,roy112,zheng12,lee15}. This method involves solving a set of coupled equations which becomes cumbersome when the quantum system is few-body or when there is a large number of incident photons. This was limiting early works to at most two-photon scattering by two noninteracting simple quantum emitters. More recently, with the use of matrix product operator, path integration, Green's function, and diagrammatic techniques \cite{pletyukhov12,caneva15,shi15,xu15,see17,das181,das182,manasi18}, multiphoton scattering studies have been extended to a few interacting emitters where signatures of strongly correlated effects such as the Mott insulator transition have been studied \cite{lee15,see17,pedersen17}.

In this work, we apply our earlier diagrammatic method \cite{see17} to probe the interplay between interaction and disorder in a few-body photonic lattice. In particular, we consider the situation where two waveguides are coupled to a photonic system described by the Bose-Hubbard Hamiltonian in the presence of a quasiperiodic potential. In this model, the system's eigenstates are expected to change from extended to localised in line with the celebrated many-body localisation (MBL) transition \cite{huse14,kondov15,schreiber15,bordia16,choi16,kaufman16,smith16}. To quantify this transition for closed systems, different quantities such as level statistics, eigenstate entanglement, and participation ratio were proposed theoretically  \cite{bardarson12,iyer13,serbyn13,huse14,kjall14,khemani17}, and some were tested in a recent experiment using superconducting circuits \cite{roushan17}. In the open optical system considered here, we show how the usual optical spectroscopy techniques based on analyzing the transmission and reflection spectra can capture finite size signatures of the corresponding localisation transition for a range of parameters. We calculate and compare the two-photon transmission probability of the open system to the participation ratio of the two-particle eigenstates of the closed system and find close agreement for a range of coupling and decay rates.

We start by reviewing the waveguide QED setup and the Hamiltonian of interest in \cref{sec:setup}. We review details of the scattering formalism required to compute the two-photon transmission probability in \cref{sec:scattering} and discuss the interpretation of two-photon transmission probability, and its role in characterizing the transition in a quantitative manner, in \cref{sec:relation}. We then compare the behavior of the transmission probability against the known behavior of the participation ratio in \cref{sec:results}. We first focus on the linear case when interaction is absent, i.e., the Aubry-Andr\'e (AA) model, to investigate the delocalisation-localisation transition or metal-to-insulator transition (MIT). After that, we show how the presence of interactions affects the two-photon transmission probability of the system. The effect of losses, which is inherent in a waveguide QED setup, is examined in \cref{sec:loss}. Finally, the consequences of using different input states are discussed in \cref{sec:coherent}. 

\section{Setup and model}\label{sec:setup}
\begin{figure}[!hbt]
\centering
\sidesubfloat[]{
\includegraphics[width=0.9\textwidth]{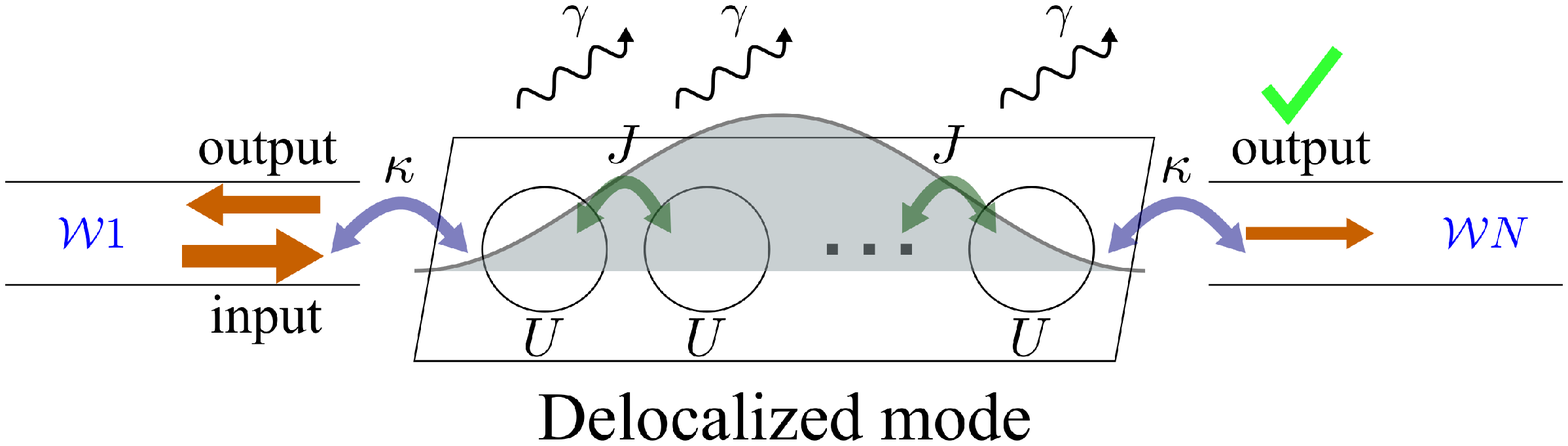}}

\vspace{1em}
\sidesubfloat[]{
\includegraphics[width=0.9\textwidth]{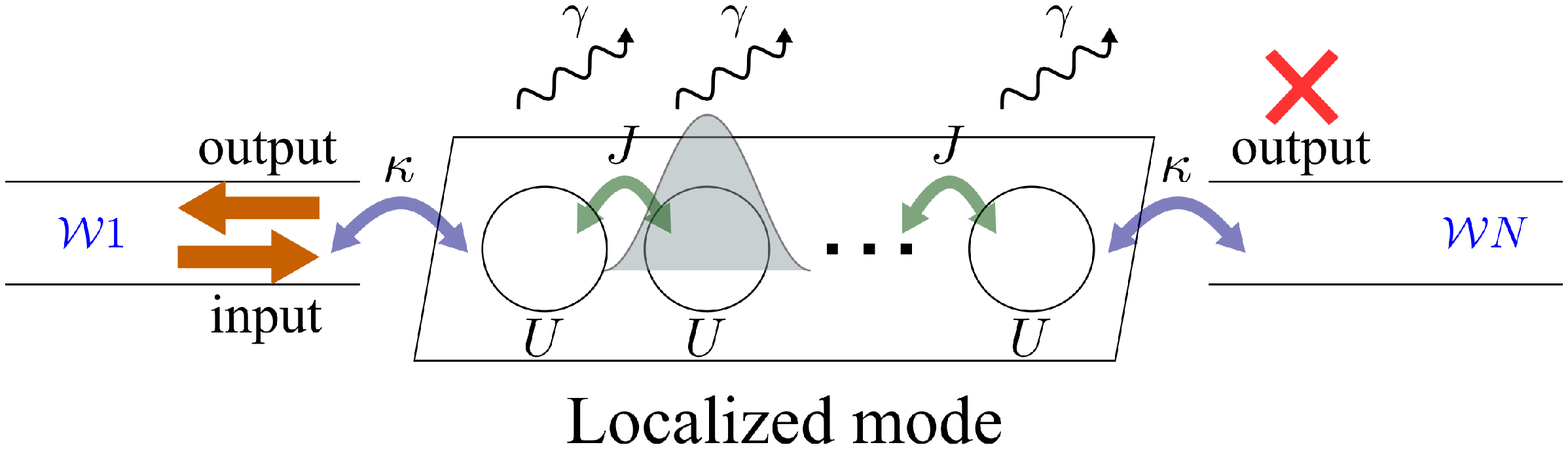}}

\vspace{1em}
\sidesubfloat[]{
\includegraphics[width=0.9\textwidth]{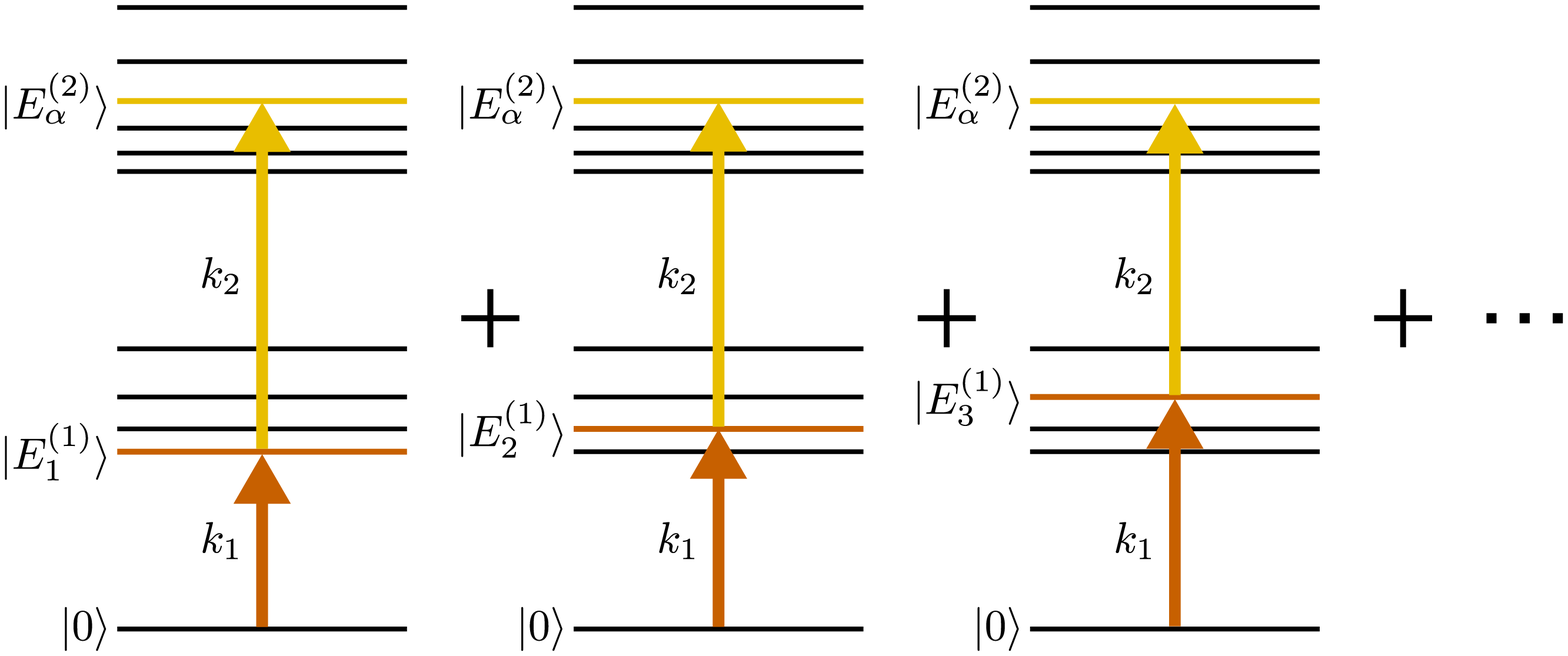}}

\caption{Two waveguides are coupled to each end of the lattice. We input two photons with momenta $k_1$ and $k_2$ from waveguide $\mathcal{W}1$ and investigate the statistics of the scattered photons at waveguide $\mathcal{W}N$ as a function of disorder for a fixed value of the onsite interaction. The sum of the momenta $k_1+k_2$ is resonant with the eigenstate $\ket{E^{(2)}_\alpha}$. We also consider how losses in the one-dimensional chain affect the statistics of the output light. (a) If $\ket{E^{(2)}_\alpha}$ is delocalised over the system, it will couple strongly to both waveguides $\mathcal{W}1$ and $\mathcal{W}N$, allowing the two input photons to be transmitted at $\mathcal{W}N$. (b) If $\ket{E^{(2)}_\alpha}$ is localised, it will couple weakly to both waveguides. Hence, the two-photon transmission at $\mathcal{W}N$ is strongly suppressed. (c) A diagram showing the choices of $k_1$ and $k_2$. To maximise the probability of the first photon entering the system, $k_1$ is set to be resonant with one of the one-particle eigenstates $\ket{E^{(1)}_\mu}$. The second photon then populates $\ket{E^{(2)}_\alpha}$ from $\ket{E^{(1)}_\mu}$. However, since we are only interested in the properties of $\ket{E^{(2)}_\alpha}$, we later reduce the contribution of $\ket{E^{(1)}_\mu}$ by varying $k_1$ to be resonant with \textit{all possible} $\ket{E^{(1)}_\mu}$, and averaging the two-photon transmission probability measured at $\mathcal{W}N$. The resonant condition for the input state also ensures that two-photon scattering has the largest contribution in the output.}
\label{fig:setup}
\end{figure}

We consider a setup as in \cref{fig:setup} where two identical waveguides are coupled to each end of a system ($\mathcal{W}1$ to site 1 and $\mathcal{W}N$ to site N) described by the Bose-Hubbard Hamiltonian with an incommensurate potential. The setup has the total Hamiltonian
\begin{equation}
\hat{H}_\text{tot} = \hat{H}_\text{w} + \hat{H}_\text{sys} + \hat{H}_\text{ws},
\end{equation}
where the first term 
\begin{equation}
\hat{H}_\text{w} = \sum_{j=1,N}\int \rmd k \; k \;\hat{b}_k^{j\dagger}\hat{b}_k^j
\end{equation}
describes the free propagating photons in both waveguides with $\hat{b}^j_k$ ($\hat{b}_k^{j\dagger}$) the photon annihilation (creation) operator in the waveguide $\mathcal{W}j$. The second term is the Hamiltonian of interest \cite{michal14} given by
\begin{multline}\label{eq:AAwint}
\hat{H}_\text{sys} = \sum_{j = 1}^N \left(\epsilon_j \hat{a}_j^\dagger \hat{a}_j + \frac{U}{2}\hat{a}_j^\dagger \hat{a}_j^\dagger \hat{a}_j \hat{a}_j\right) \\ + J \sum_{j = 1}^{N - 1} \left(\hat{a}_j^\dagger \hat{a}_{j + 1} + \text{H.c.}\right),
\end{multline}
where $\hat{a}_j$ ($\hat{a}_j^\dagger$) is the annihilation (creation) operator at site $j$ with $[\hat{a}_i, \hat{a}_j^\dagger] = \delta_{ij}$, and $\epsilon_j = h\cos(2\pi bj)$ is the on-site energy. $J$ and $U$ are the hopping and interaction strengths, respectively. The last term, 
\begin{equation}\label{eq:coupling}
\hat{H}_\text{ws} = \sum_{j=1,N}\int \rmd k\sqrt{\frac{\kappa}{2\pi}}\left(\hat{b}_k^{j\dagger}a_j + \text{H.c.}\right),
\end{equation}
describes the system-waveguide interaction.

The Hamiltonian in \cref{eq:AAwint}, $\hat{H}_\text{sys}$ preserves the total number of particles $\hat{N}=\sum_{j = 1}^N  \hat{a}_j^\dagger \hat{a}_j $. Hence, from now on, we will denote the $d_M = {N + M - 1\choose M}$ eigenstates in the $M$-particle manifold by $\ket{E_\alpha^{(M)}}$ with eigenenergies $E_\alpha^{(M)}$ for $\alpha = 1, 2, \dots, d_M$ and $E_1^{(M)} \leq E_2^{(M)} \leq \dots \leq E_{d_M}^{(M)}$. 

When $U = 0$, \cref{eq:AAwint} is equivalent to the Aubry-Andr\'e (AA) model. As shown in \cite{aubry80}, when the potential is incommensurate, i.e., $b = \frac{\sqrt{5} - 1}{2}$, the model exhibits a MIT at the critical potential strength $h/J = 2$.  If one takes into account the effects of interaction between particles, the situation can change dramatically. For example, it has been shown that there exists delocalised two-particle bound states \cite{flach12,frahm15} in the localised phase of the noninteracting AA model (when $h/J > 2$). It has also been shown that such a MIT becomes an ergodic-MBL transition in the presence of interactions between the particles \cite{iyer13}. In this case, the localisation properties of the states depend on the energy, and one can possibly have a many-body mobility edge, though its existence is still under debate \cite{deroeck16}. Recently, the AA model with interaction has been implemented using superconducting circuits \cite{roushan17}. 

\subsection{Photon scattering for few-body spectroscopy}\label{sec:scattering}
In order to observe interaction-induced effects, at least two photons are required to be in the system. For this, we send two photons to the system, each with different momenta $k_1$ and $k_2$ via waveguide $\mathcal{W}1$, denoted by the input state $\ket{\text{in};k_1, k_2}$. The probability of detecting two photons in the waveguide $\mathcal{W}N$, i.e., the two-photon transmission probability, for a given input $\ket{\text{in};k_1, k_2}$, is
\begin{equation}\label{eq:prob}
P(k_1, k_2) := \frac{1}{2}\int \rmd p_1 \rmd p_2 \;\rho(p_1, p_2|k_1, k_2),
\end{equation}
where $\rho(p_1, p_2|k_1, k_2)$ is the conditional probability of detecting two photons with momenta $p_1$ and $p_2$ via waveguides $\mathcal{W}N$. It has the expression
\begin{align}\label{eq:rho}
\rho(p_1, p_2|k_1, k_2) = &\bigg\lvert\braket{0|\hat{b}_{p_1}^{N}\hat{b}_{p_2}^{N}|\text{out};k_1, k_2}\bigg\rvert^2 \notag\\
= &\bigg\lvert\braket{0|\hat{b}_{p_1}^{N}\hat{b}_{p_2}^{N}\hat{S}|\text{in};k_1, k_2}\bigg\rvert^2,
\end{align}
with $\hat{S}$ the scattering operator and $\ket{\text{out};k_1, k_2} = \hat{S}\ket{\text{in};k_1, k_2}$.

To investigate the properties of a particular two-particle eigenstate, $\ket{E_\alpha^{(2)}}$ of \cref{eq:AAwint}, we consider input photons with momenta $k_1$ and $k_2$ such that $k_1 + k_2 = E_\alpha^{(2)}$. Moreover, we want to consider cases where two-photon scatterings are fully resonant, i.e., one of the photons has to resonantly excite one of the single-particle eigenstates, $\ket{E_\mu^{(1)}}$. The resonant condition reads
\begin{equation}\label{eq:resonant}
k_1=E_\mu^{(1)}, \; k_2 = E_\alpha^{(2)}-E_\mu^{(1)} \qquad\text{with}\qquad \mu = 1, \dots, d_1.
\end{equation}
In order to dilute the effects of a particular single-particle eigenstate, we take an average over all $d_1$ resonant paths, and define the quantity
\begin{equation}\label{eq:T2}
T^{(2)}(\alpha) := \frac{1}{d_1}\sum_{\mu = 1}^{d_1} P\left(E^{(1)}_\mu, E^{(2)}_\alpha-E^{(1)}_\mu\right).
\end{equation}
$T^{(2)}(\alpha)$ is simply the two-photon transmission probability of a given eigenstate $\ket{E_\alpha^{(2)}}$ averaged over all paths where the two-photon transitions are resonant. The input state is chosen such that the contribution arising from two-photon scattering into the desired eigenstate is the largest at the output. The intuition behind the choice of this quantity is discussed in \cref{fig:setup}. 

Next, we investigate how $T^{(2)}(\alpha)$ is able to show signatures of a delocalisation-localisation transition by comparing it with the known behavior of the participation ratio \cite{kramer93}. The participation ratio is defined as
\begin{equation}\label{eq:PR}
R(\alpha) := \frac{1}{\sum_{i \leq j} |c^\alpha_{i, j}|^4}
\end{equation}
where $\ket{E_\alpha^{(2)}}=\sum_{i \leq j} c^\alpha_{i, j}\ket{i, j}$ and  $c^\alpha_{i, j} = \braket{i, j|E_\alpha^{(2)}}$ are the coefficients of $\ket{E_\alpha^{(2)}}$ in the Fock state basis $\{\ket{i, j} = \hat{a}_i^\dagger\hat{a}_j^\dagger/\sqrt{1 + \delta_{ij}}\ket{0} ; i \leq j\}$. When an eigenstate is delocalised, $c_{i, j}^\alpha \approx 1/\sqrt{d_2}$ for all $(i, j)$ and hence $R(\alpha) \approx d_2$ and $\log_{d_2}R(\alpha) \approx 1$; however, when an eigenstate is localised around a configuration $\ket{i_0, j_0}$, $c_{i_0, j_0}^\alpha \approx 1$ and $c_{i, j}^\alpha \approx 0$ otherwise, hence $R(\alpha) \approx 1$ and $\log_{d_2}R(\alpha) \approx 0$. 

The participation ratio, $R(\alpha)$ can be computed by diagonalizing the Hamiltonian whereas the two-photon transmission probability $T^{(2)}(\alpha)$ is fully characterised by the scattering operator. These quantities are intimately related. The participation ratio $R(\alpha)$ is a measure of how extended a given eigenstate  $\ket{E_\alpha^{(2)}}$ is in the Fock state basis $\{\ket{i, j}\}$.  Certainly, transmission is possible when the eigenstates are delocalised because there are overlaps between the states and the waveguides $\mathcal{W}1$ and $\mathcal{W}N$ at sites $1$ and $N$ respectively. Conversely, if a given eigenstate with energy $E_\alpha^{(2)}$ is localised in the Fock state basis, the probability that two photons can be transmitted between the waveguides is small. 

In the following subsection, we will outline the steps to compute the scattering elements, which allows one to reconstruct the scattering operator. After which, we provide a quantitative analysis on the motivation for proposing the quantity $T^{(2)}$, and discuss the role the resonant condition \cref{eq:resonant} plays in making $T^{(2)}$ a good diagnostic tool for the delocalisation-localisation transition. We do this before the discussion of the results in order to contrast the behavior of the two-photon transmission probability $T^{(2)}(\alpha)$ with the logarithm of the participation ratio $\log_{d_2}R(\alpha)$. Both quantities are expected to vary from a finite value towards zero as $h/J$ increases, because the system undergoes a delocalisation-localisation transition around $h/J = 2$. We would like to emphasise that in the interacting case, the localisation length of the states depends on the energy. Therefore, we expect $T^{(2)}(\alpha)$ and $\log_{d_2}R(\alpha)$ to have a dependency on $E_\alpha^{(2)}$. 

\subsection{Transmission spectra from scattering theory}\label{sec:smatrix}
Following the formalism developed in \cite{xu15}, the scattering elements of this setup can be calculated using the effective Hamiltonian 
\begin{equation}\label{eq:Heff}
\hat{H}_\text{eff} = \hat{H}_\text{sys} - \rmi\frac{\kappa}{2}\left(\hat{a}_1^\dagger \hat{a}_1 + \hat{a}_N^\dagger \hat{a}_N\right),
\end{equation} 
instead of the full Hamiltonian.
Note that $\hat{H}_\text{eff}$ is non-Hermitian and therefore its eigenenergies are complex in general. From now on, $\ket{\xi_{\alpha}^{(M)}}$ and $\bra{\bar{\xi}_{\alpha}^{(M)}}$ will denote the right and left $M$-particle eigenstates of $\hat{H}_\text{eff}$ with eigenenergies $\xi_{\alpha}^{(M)}$. As before, there are $d_M$ eigenstates in the $M$-particle manifold of $\hat{H}_\text{eff}$. 

With this, we have all the elements that we need for the diagrammatic approach in \cite{see17}. The advantage of the latter is that the expressions for the scattering elements can be written down directly with the aid of scattering diagrams. Our aim in this section is to explicitly calculate the probability, $P(k_1, k_2)$ by integrating out $\rho(p_1, p_2|k_1, k_2)$ as in \cref{eq:prob}. The probability, $P(k_1, k_2)$ is of utmost importance because it allows us to obtain the two-photon transmission $T^{(2)}(\alpha)$ in \cref{eq:T2}.

We consider input states with a finite momentum width by defining the operator $\hat{B}_k^{1\dagger} := \int \rmd q \;\chi_k(q) \hat{b}_q^{1\dagger}$, where $\chi_k(q) = \sqrt{\frac{\sigma}{\pi}}\frac{1}{q - k + \rmi\sigma}$. This creates a photon in the waveguide $\mathcal{W}1$ with a Lorentzian momentum profile centred around $k$ with width $\sigma$ (we set $\sigma = 0.01J$ in this paper). Thus, the input state is $\ket{\text{in};k_1, k_2} = \frac{1}{\sqrt{M(k_1, k_2)}}\hat{B}_{k_1}^{1\dagger}\hat{B}_{k_2}^{1\dagger}\ket{0}$, where $M(k_1, k_2) := 1 + 4\sigma^2/((k_1 - k_2)^2 + 4\sigma^2)$ is the normalisation constant. The conditional probability of detecting two photons of momenta $p_1$ and $p_2$ at $\mathcal{W}N$, $\rho(p_1, p_2|k_1, k_2)$ defined in \cref{eq:rho}, is then given by
\begin{align}\label{eq:rhos}
&\rho(p_1, p_2|k_1, k_2) \notag \\
= &\frac{1}{M(k_1, k_2)}\bigg\lvert\int \rmd q_1 \rmd q_2 \;\chi_{k_1}(q_1)\chi_{k_2}(q_2)S(p_1, p_2; q_1, q_2)\bigg\rvert^2.
\end{align}
In the previous expression, $S(p_1,p_2;q_1,q_2)$ is the two-photon S-matrix element with input photons of momenta $q_1$ and $q_2$ via $\mathcal{W}1$ and output photons of momenta $p_1$ and $p_2$ via $\mathcal{W}N$ \footnote{In general, there are other S-matrix elements with input and output from all possible combinations of $\mathcal{W}1$ and $\mathcal{W}N$, but we will only be considering the aforementioned S-matrix element.}.

From the previous discussion, we can see that the core of the calculation is to find the two-photon S-matrix element, $S(p_1, p_2; q_1, q_2)$. The latter can be decomposed in terms of the one-photon S-matrix element $S(p; q)$ and the four-point Green's function, i.e.
\begin{multline}
S(p_1, p_2; q_1, q_2) = S(p_1;q_1)S(p_2;q_2) \\+ S(p_2;q_1)S(p_1;q_2) + G(p_1, p_2; q_1, q_2)
\end{multline}
where the one-photon S-matrix element equals the two-point Green's function \footnote{If the input and output channels are the same, i.e. in the case of reflection, the S-matrix element is equal to the corresponding 2-point Green's function plus a delta function.},
\begin{equation}
S(p; q) = G(p;q).
\end{equation}

\begin{figure}[!h]
\sidesubfloat[]{
\includegraphics[width=0.35\textwidth]{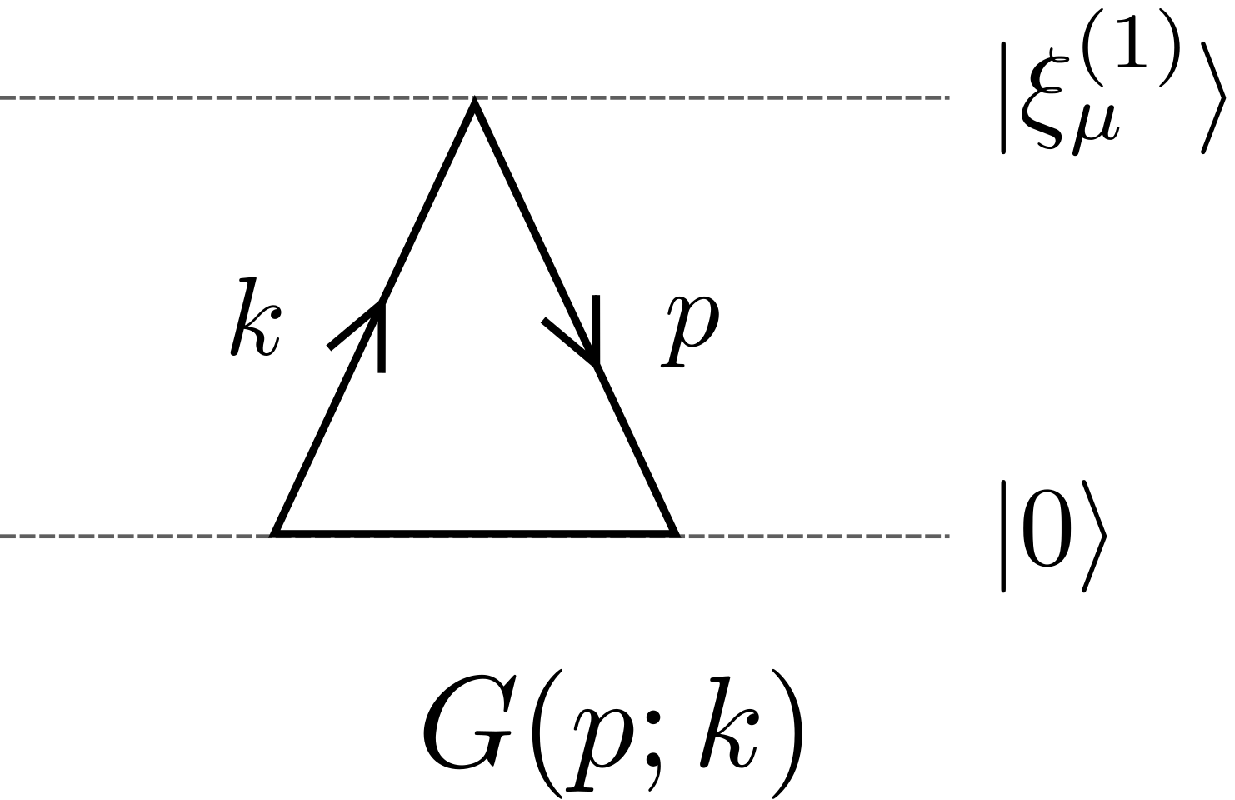}}\label{fig:scattd2}

\vspace{1em}
\sidesubfloat[]{
\includegraphics[width=0.85\textwidth]{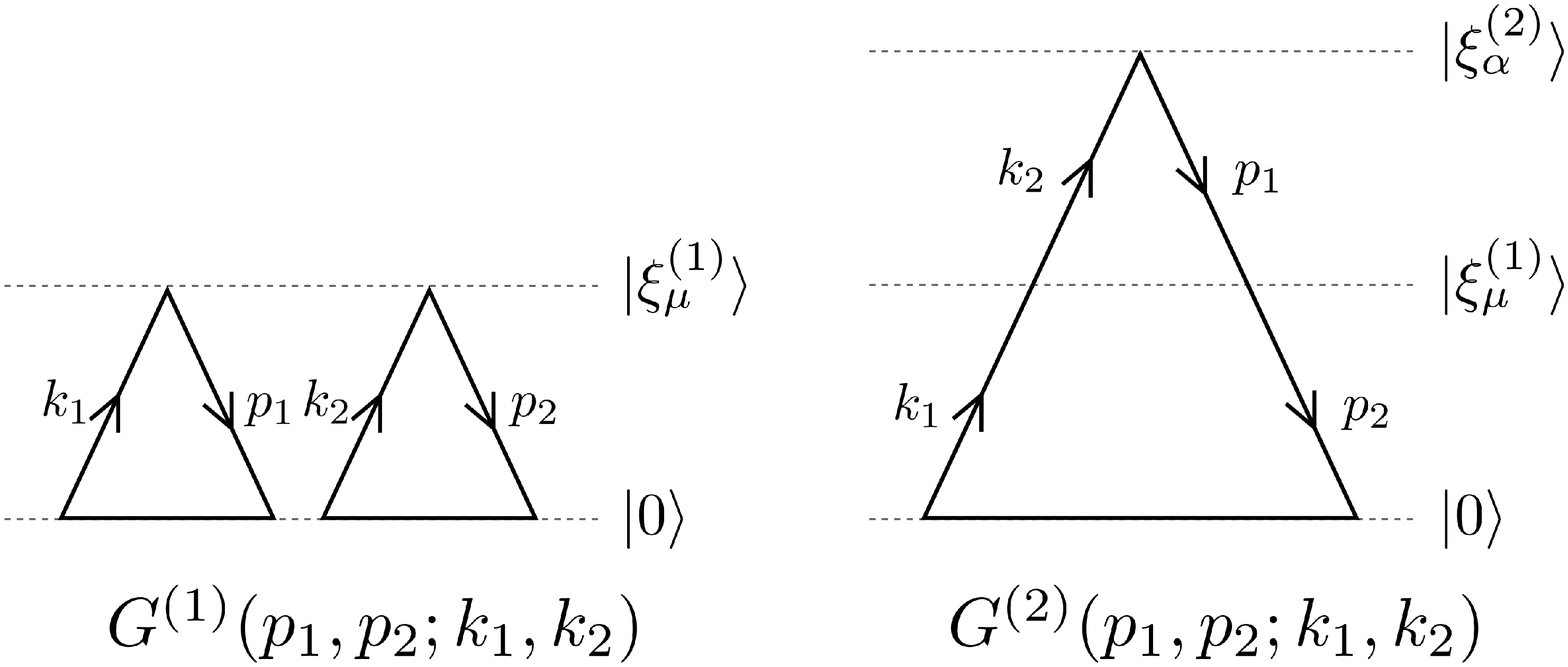}}\label{fig:scattd4}

\caption{Diagrammatic method to calculate the scattering elements by summing up all possible optical absorption and emission paths. Scattering diagrams for a) two-point Green's function and b) four-point Green's function.}\label{fig:scattd}
\end{figure}

Here is where we can appreciate the power of our diagrammatic approach \cite{see17}: the two- and four-point Green's functions can be represented by scattering diagrams, as depicted in \cref{fig:scattd} showing all possible optical absorption and emission paths. The expressions for the Green's functions can then be written down directly based on the diagrams. For example, the two-point Green's function is given by
\begin{equation}\label{eq:2point}
G(p;q) = \sum^{d_1}_{\mu=1}\braket{0|\hat{a}_N|\xi_{\mu}^{(1)}}\braket{\bar{\xi}_{\mu}^{(1)}|\hat{a}^\dagger_1|0}\frac{-\rmi\kappa}{q - \xi_{\mu}^{(1)}}\delta(p - q)
\end{equation}
with the system operators $\hat{a}_N$ and $\hat{a}_1^\dagger$ evolving according to the effective Hamiltonian in \cref{eq:Heff} and the sum is taken over all $d_1$ single-particle eigenenergies, $\xi_\mu^{(1)}$, of $\hat{H}_\text{eff}$. Similarly, the four-point Green's function,
\begin{equation}\label{eq:4point}
G(p_1, p_2; q_1, q_2) = \sum_{l = 1}^2 G^{(l)}(p_1, p_2; q_1, q_2)
\end{equation}
with
\begin{align}
&G^{(1)}(p_1, p_2; q_1, q_2) \notag \\
= &-\frac{\rmi\kappa^2}{2\pi}\delta(p_1 + p_2 - q_1 - q_2) \notag \\
&\times\sum_{\mu, \nu}\Bigg(\frac{\braket{0|\hat{a}_{N}|\xi_{\nu}^{(1)}}\braket{\bar{\xi}_{\nu}^{(1)}|\hat{a}^\dagger_{1}|0}\braket{0|\hat{a}_{N}|\xi^{(1)}_{\mu}}\braket{\bar{\xi}^{(1)}_{\mu}|\hat{a}^\dagger_{1}|0}}{\left(p_2 - \xi^{(1)}_{\nu}\right) \left(q^{\vphantom{(1)}}_1 - p^{\vphantom{(1)}}_1\right) \left(q_1 - \xi^{(1)}_{\mu}\right)} \notag\\
&+ \;\text{all permutations of $\{q_1, q_2\}$ and $\{p_1, p_2\}$}\Bigg) \label{eq:twog1}\\
\intertext{and}
&G^{(2)}(p_1, p_2; q_1, q_2) \notag \\
= &-\frac{\rmi\kappa^2}{2\pi}\delta(p_1 + p_2 - q_1 - q_2) \notag\\
&\times\sum_{\mu, \beta, \nu}\Bigg(\frac{\braket{0|\hat{a}_{N}|\xi^{(1)}_{\nu}}\braket{\bar{\xi}^{(1)}_{\nu}|\hat{a}_{N}|\xi^{(2)}_{\beta}}\braket{\bar{\xi}^{(2)}_{\beta}|\hat{a}^\dagger_{1}|\xi^{(1)}_{\mu}}\braket{\bar{\xi}^{(1)}_{\mu}|\hat{a}^\dagger_{1}|0}}{\left(p_2 - \xi^{(1)}_{\nu}\right)\left(q_1 + q_2 - \xi^{(2)}_{\beta}\right)\left(q_1 - \xi^{(1)}_{\mu}\right)}\notag\\
&+ \;\text{all permutations of $\{q_1, q_2\}$ and $\{p_1, p_2\}$}\Bigg). \label{eq:twog2}
\end{align}
Again, the system operators $\hat{a}_N$ and $\hat{a}_1^\dagger$ evolve according to the effective Hamiltonian, $\hat{H}_\text{eff}$ and the sums are taken over all single-particle eigenenergies, $\xi^{(1)}_{\mu}$ and $\xi^{(1)}_{\nu}$ and two-particle eigenenergies, $\xi^{(2)}_{\beta}$ of $\hat{H}_\text{eff}$. With the knowledge of $S(p_1,p_2; q_1, q_2)$, the probability $P(k_1, k_2)$ is then calculated by doing the integration in \cref{eq:rhos,eq:prob}.

\subsection{Relationship between transmission probability and participation ratio}\label{sec:relation}
In this section, we will discuss the relationship between the two-photon transmission probability, $T^{(2)}$, and the participation ratio, $R$, by looking at the resolvent operator $\hat{\mathcal{G}}(E) := \left(E - \hat{H}_\text{sys}\right)^{-1}$. In \cite{vonoppen96}, the authors defined the localisation length of two-interacting particles (TIP) as
\begin{equation}\label{eq:ll2}
\frac{1}{\lambda_2} = - \lim_{N\rightarrow\infty} \frac{1}{2(N - 1)}\ln\big|\braket{N, N|\hat{\mathcal{G}}|1, 1}\big|^2,
\end{equation}
where $\ket{i, j} = \hat{a}_i^\dagger\hat{a}_j^\dagger/\sqrt{1 + \delta_{ij}}\ket{0}$  with $i \leq j$ is the Fock state defined previously. This definition is a direct generalisation of the linear case where the single-particle localisation length is defined as \cite{kramer93} 
\begin{equation}\label{eq:ll1}
\frac{1}{\lambda_1} = - \lim_{N\rightarrow\infty} \frac{1}{2(N - 1)}\ln\big|\braket{N|\hat{\mathcal{G}}|1}\big|^2.
\end{equation}

Here, we show that for TIP, the two-photon transmission probability $T^{(2)}$ defined in \cref{eq:T2} is approximately proportional to the matrix element in the definition of TIP's localisation length, i.e.
\begin{equation}\label{eq:ReseqT2}
T^{(2)} \approx \mathcal{C}|\braket{N, N|\hat{\mathcal{G}}|1, 1}|^2, 
\end{equation}
where $\mathcal{C}$ is a constant. For $U = 0$, the four-point Green's function vanishes and one can evaluate $T^{(2)}$ easily and compare it with $|\braket{N, N|\hat{\mathcal{G}}|1, 1}|^2$ expressed in spectral representation. For $U \neq 0$, note that $T^{(2)}$ is defined for very specific input states, namely states that satisfy the resonant condition \cref{eq:resonant}. The resonant condition \cref{eq:resonant} guarantees that the two-photon scattering events are fully resonant and thus have the largest contribution to the transmission probability. In other words, $T^{(2)}$ will consist mainly of contribution from fully resonant two-photon scattering processes corresponding to \cref{eq:twog2}. The other terms, \cref{eq:2point,eq:twog1}, in its expression correspond to off-resonant processes and hence have small amplitudes. The argument is laid out in more detail in \cref{sec:analytic}. Using \cref{eq:ReseqT2}, one can estimate the TIP's localisation length using $T^{(2)}$ by 
\begin{equation}\label{eq:Lambda2}
\frac{1}{\lambda_2} \approx \frac{1}{\Lambda_2} := - \lim_{N\rightarrow\infty} \frac{1}{2(N - 1)}\ln T^{(2)},
\end{equation}
where we have defined the quantity $\Lambda_2$ as an effective TIP's localisation length derived from the two-photon transmission probability. 

Following the line of reasoning above, it is clear that when $U \neq 0$ the connection between $T^{(2)}$ and $|\braket{N, N|\hat{\mathcal{G}}|1, 1}|^2$ relies heavily on the dominance of two-photon resonant paths, achieved by choosing appropriate input states. This justifies the choice of the input states in the definition of $T^{(2)}$ and one can see how the argument will fail if different input states are chosen instead, where the term describing two-photon scattering events (\cref{eq:twog2}) that is closely related to the resolvent operator's matrix element no longer has the largest contribution.

On the other hand, the participation ratio is approximately inversely proportional to the sum of the squared of the diagonal elements of the TIP's resolvent operator,
\begin{equation}\label{eq:ReseqPR-1}
R^{-1}(\alpha) = \sum_{i \leq j}\left|\braket{i, j|\hat{\mathcal{G}}(E^{(2)}_\alpha)|i, j}\right|^2,
\end{equation}
where the sum is taken over all $d_2$ two-particle Fock states. 

Quantitatively, $T^{(2)}$ is very different from $R$. However, both quantities are intrinsically related to the delocalisation-localisation transition. Since $T^{(2)}$ can be used as an estimate for TIP's localisation length, it characterises the asymptotic behavior of an eigenstate. $R$ on the other hand, corresponds to the volume occupied by an eigenstate. Therefore, one would expect both quantities to behave similarly as the system transitions from delocalisation to localisation. In the upcoming section, we show that they indeed exhibit qualitatively good agreement in the characterisation of the transition by considering different strengths of quasiperiodicity and interaction. We also provide numerical evidence in \cref{sec:coherent} on how input states which do not satisfy \cref{eq:resonant} are unable to correctly map out the delocalised/localised properties of an eigenstate.

\section{Signatures of localisation transition in the transmission spectra}\label{sec:results}
\subsection{Localisation due to quasiperiodic potential in the absence of interaction}
\begin{figure}[!htb]
\centering
\sidesubfloat[]{
\includegraphics[scale=0.35]{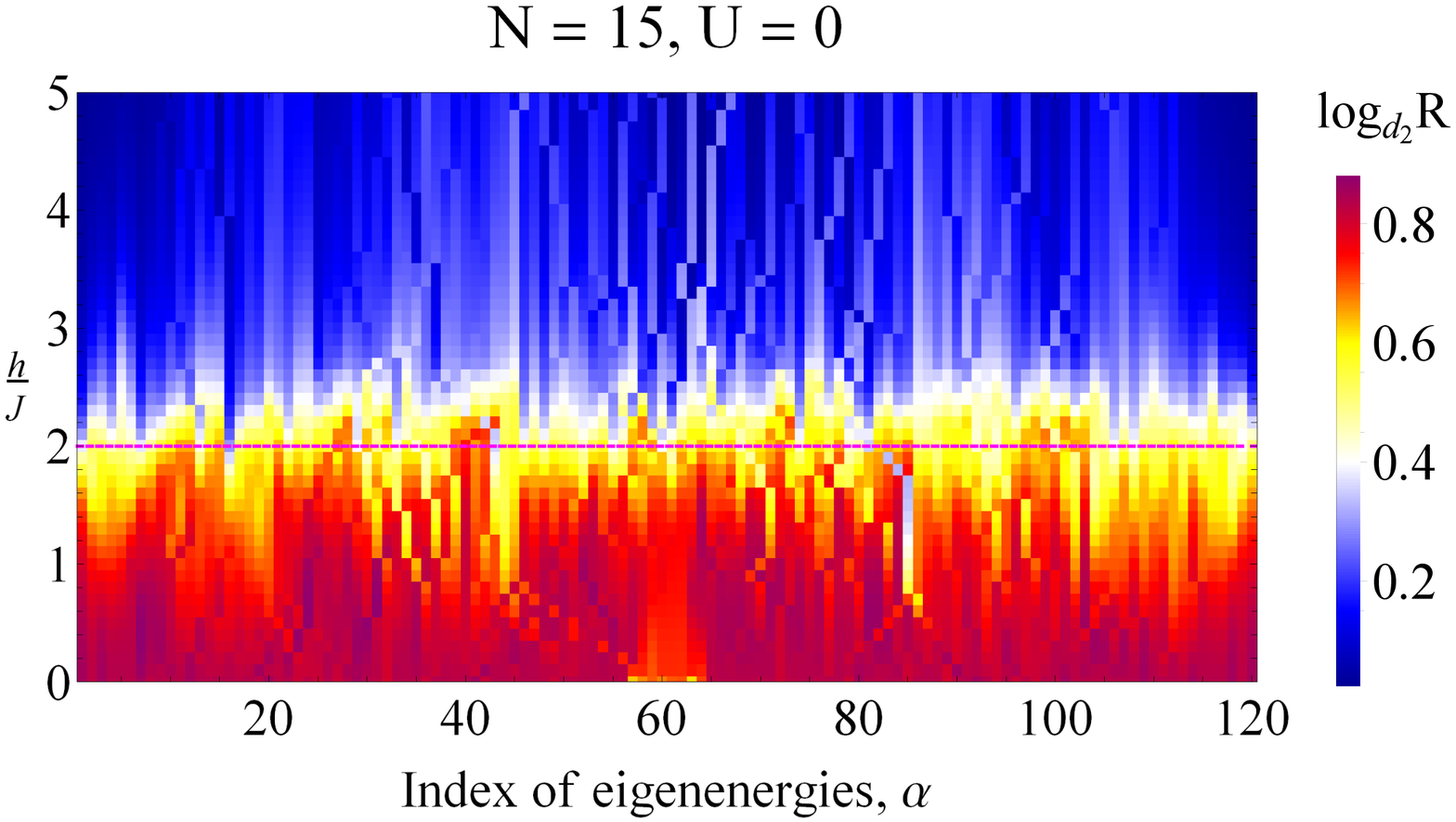}}

\vspace{1em}
\sidesubfloat[]{
\includegraphics[scale=0.35]{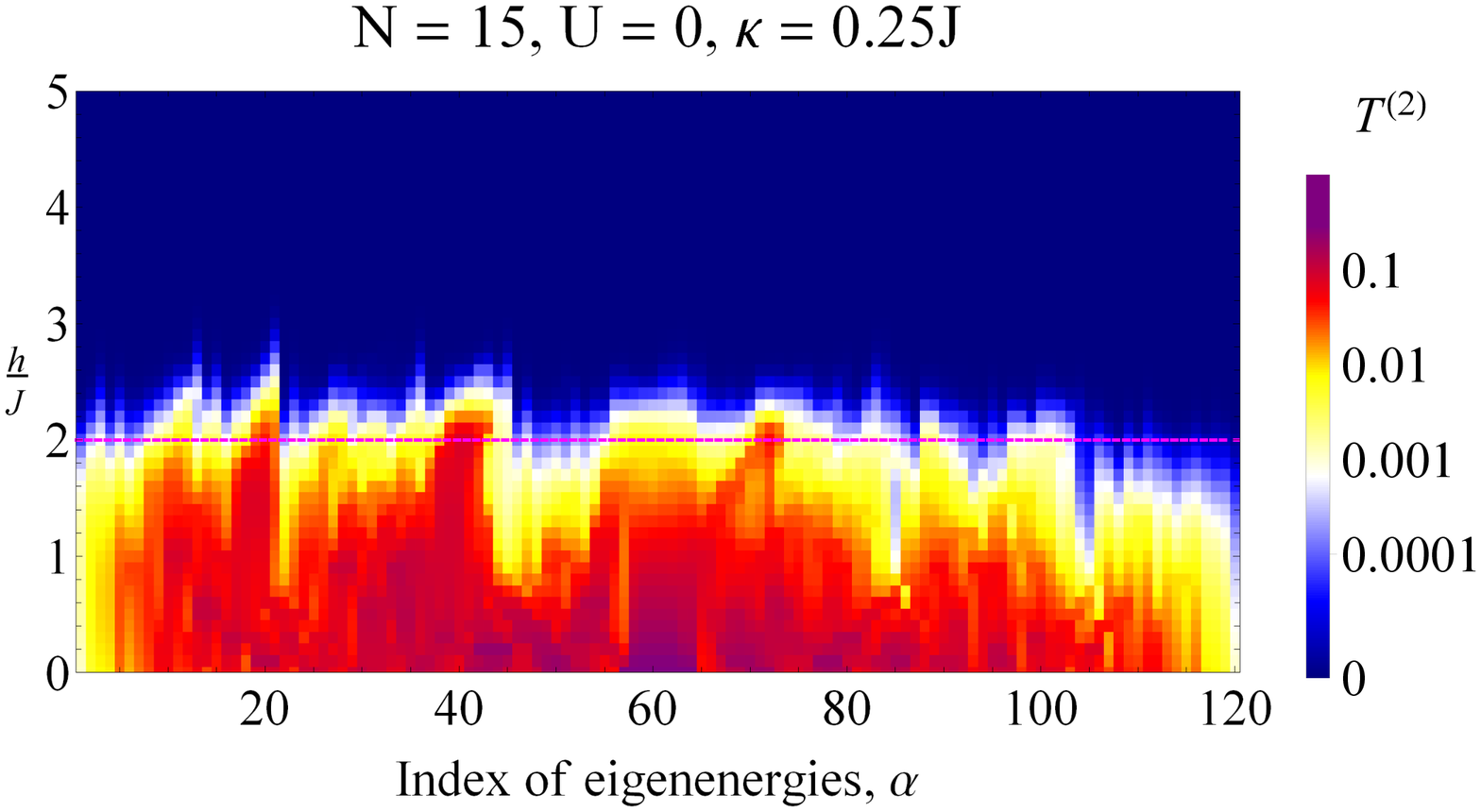}}

\caption{$U = 0$: Participation ratio vs transmission spectra. (a) The participation ratio $\log_{d_2}R(\alpha)$ and (b) the two photon transmission probability $T^{(2)}(\alpha)$, for $U = 0$. Here, $N = 15$ and $b = \frac{\sqrt{5} - 1}{2}$. Both diagrams show a transition from finite value to zero at $h/J = 2$, with the latter being an open system with a realistic waveguide system coupling of $\kappa = 0.25J$.}\label{fig:AA}
\end{figure}

In this section, we apply the methods discussed in the previous section to calculate the two-photon transmission probability. This will allow us to unveil signatures of localisation of interacting photons. For all the results in this paper, we will be using a system-waveguide coupling strength of $\kappa = 0.25J$. The rationale behind this choice is that if $\kappa/J$ is too small, the two-photon transmission will be weak. If $\kappa/J$ is too big, the probe will not be able to resolve each eigenstate, which goes against our intention of studying how interaction and quasiperiodicity affect different eigenstates. Hence, we have chosen the coupling strength such that it is smaller than the typical separation of the energy levels, which is equal to $J$ but large enough such that there is sufficient transmission. 

We first study how the two-photon transmission probability $T^{(2)}(\alpha)$ changes as $h/J$ is varied when $U = 0$, i.e., the AA model. As mentioned previously, the AA model exhibits an MIT when $h/J = 2$. This is evident in \cref{fig:AA}, which shows how the participation ratio and the two-photon transmission probability change as $h/J$ is varied for $N = 15$ lattice sites. In \cref{fig:AA}, density plots of $T^{(2)}(\alpha)$ and $\log_{d_2}R(\alpha)$ show similar behavior where they are close to zero in the localised phase ($h/J > 2$) but are nonzero while in the metallic phase ($h/J < 2$). For both quantities, the transition happens roughly at $h/J = 2$ for all eigenstates. This is no surprise since, in the linear case, the two-photon transmission probability is just the product of one-photon transmission probabilities, which are directly related to the localisation length \cite{kramer93}. 

\subsection{Localisation due to competition between quasiperiodicity and interaction}
\begin{figure}[!h]
\centering
\sidesubfloat[]{
\includegraphics[scale=0.35]{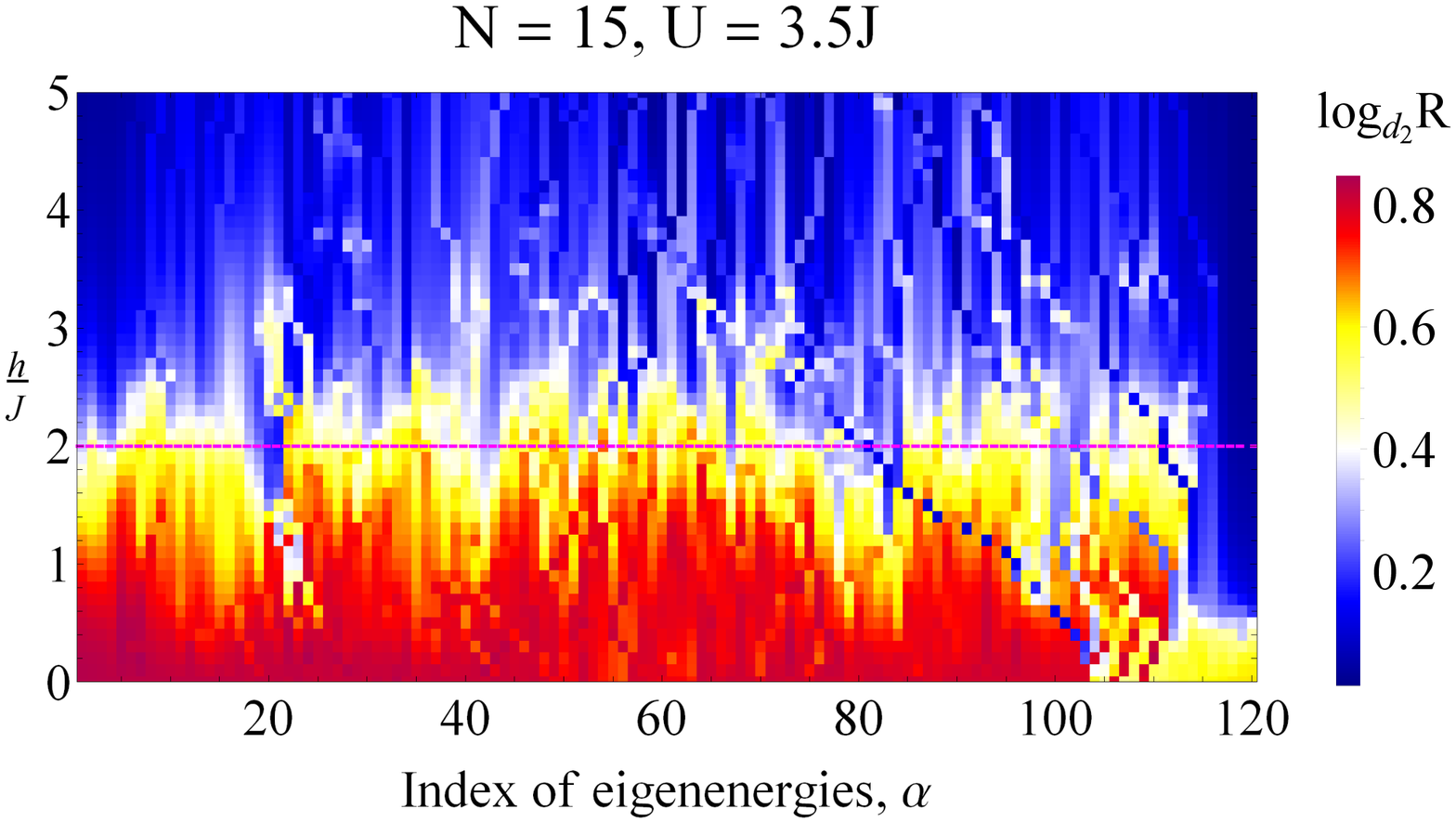}}

\vspace{1em}
\sidesubfloat[]{
\includegraphics[scale=0.35]{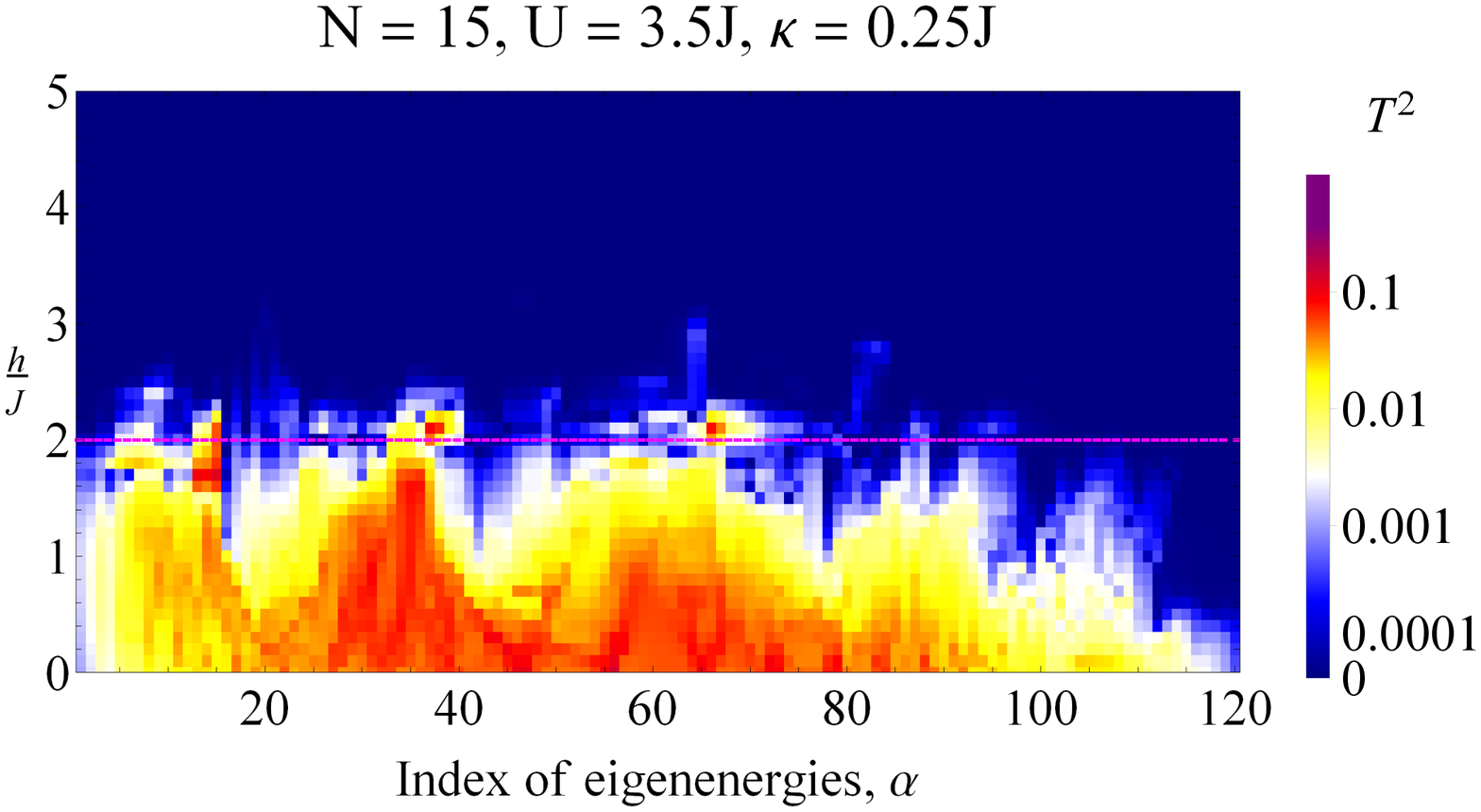}}

\caption{$U = 3.5J$: Participation ratio vs transmission spectra. Comparison between (a) the participation ratio $\log_{d_2}R(\alpha)$ and (b) the two-photon transmission probability $T^{(2)}(\alpha)$, for $U = 3.5J$. Here, $N = 15$ and $b = \frac{\sqrt{5} - 1}{2}$. For (b), $\kappa = 0.25J$. Both quantities show similar behavior. The delocalisation-localisation transition happens at different values of $h/J$ for different eigenstates.}\label{fig:n15u3.5}
\end{figure}

\begin{figure}[!h]
\centering
\sidesubfloat[]{
\includegraphics[scale=0.35]{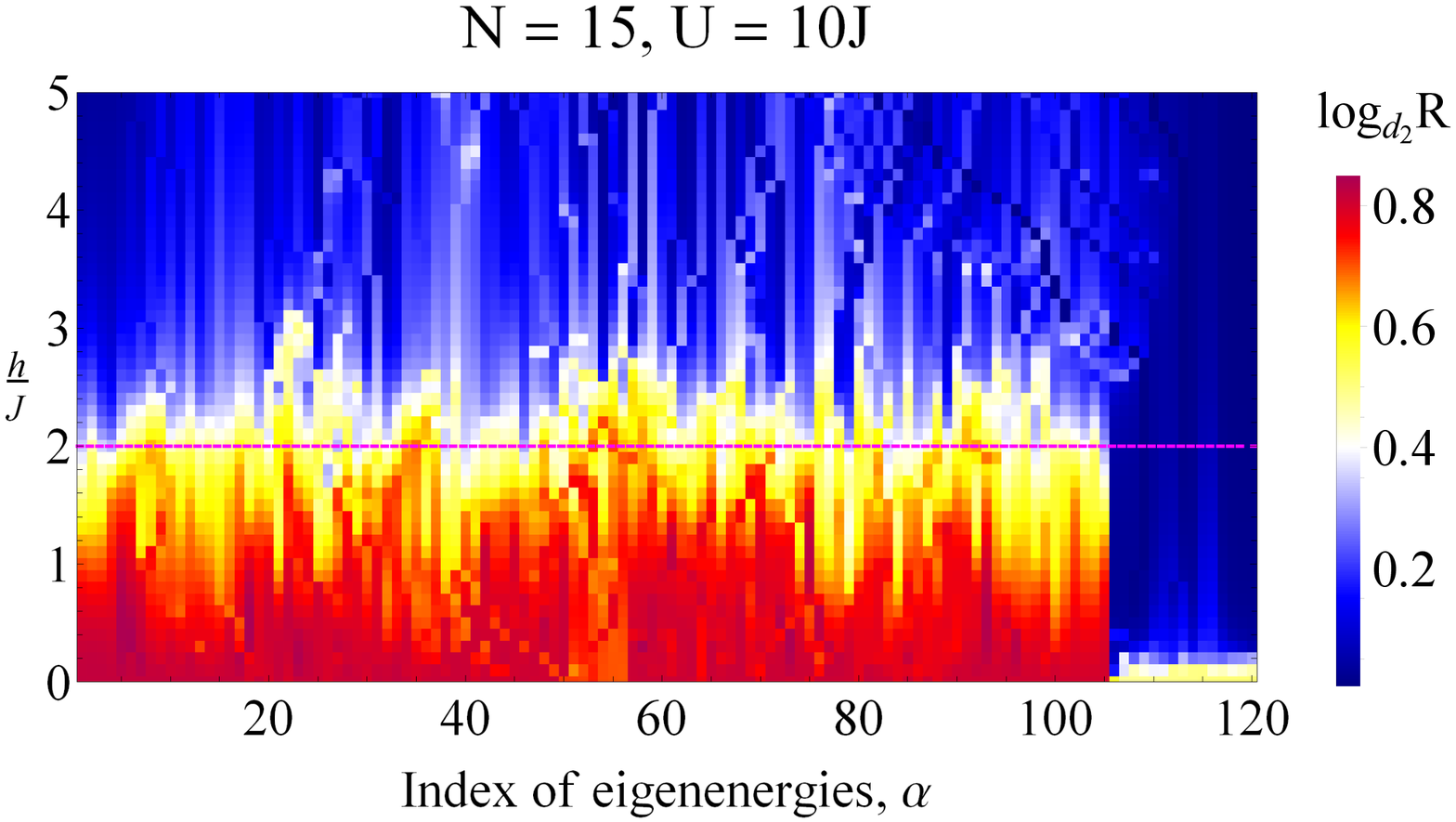}}

\vspace{1em}
\sidesubfloat[]{
\includegraphics[scale=0.35]{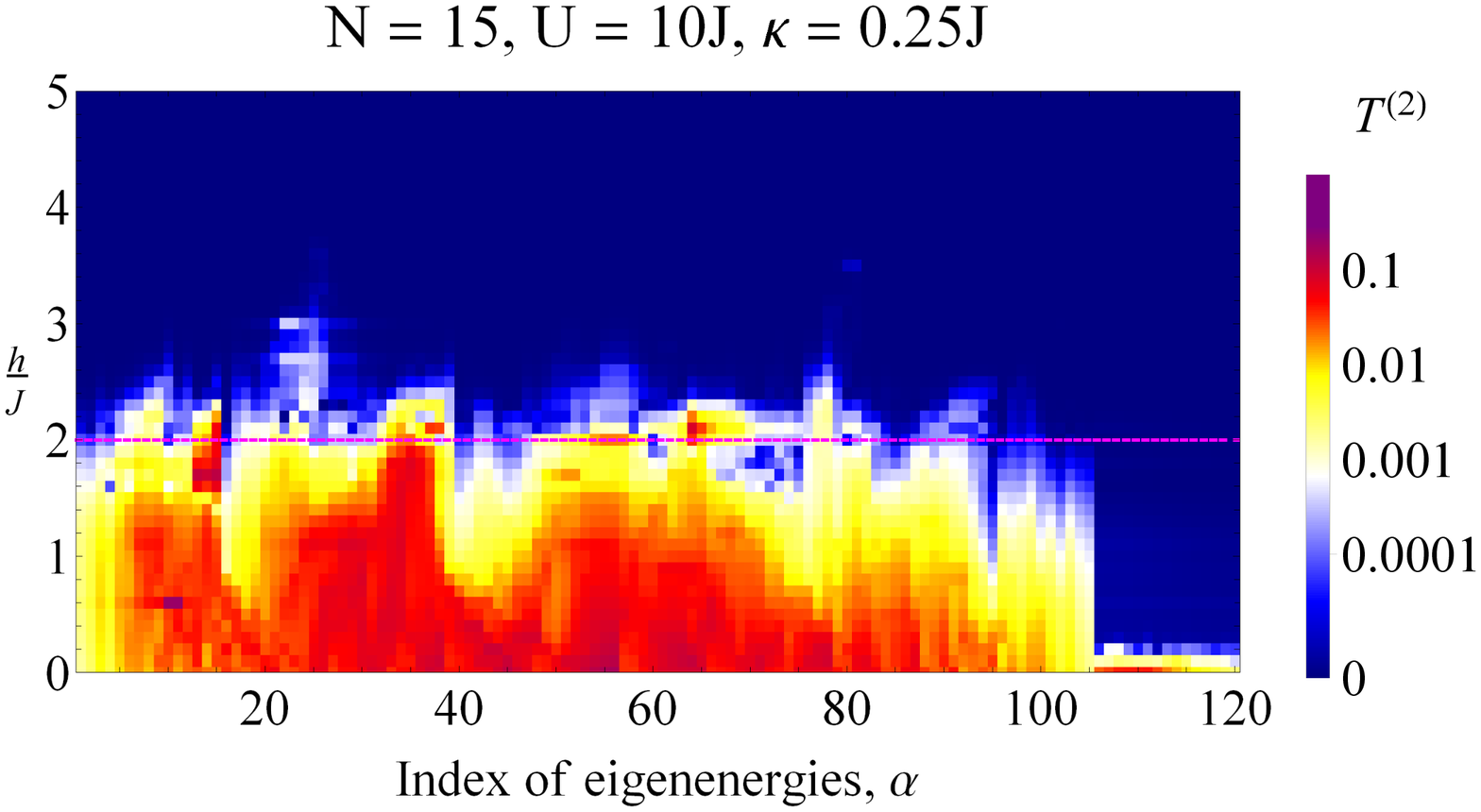}}

\caption{$U = 10J$: Participation ratio vs transmission spectra. Comparison between (a) the participation ratio $\log_{d_2}R(\alpha)$ and (b) the two-photon transmission probability $T^{(2)}(\alpha)$, for $U = 10J$. Here, $N = 15$ and $b = \frac{\sqrt{5} - 1}{2}$. For (b), $\kappa = 0.25J$. Both quantities show similar behavior. The delocalisation-localisation transition happens at different values of $h/J$ for different eigenstates.}\label{fig:n15u10}
\end{figure}

Next, we study how on-site interaction ($U\neq 0$) changes the transition. Again we will be looking at the same quantity $T^{(2)}(\alpha)$ as defined in \cref{eq:T2} and comparing it with the known behavior of the participation ratio, $\log_{d_2}R(\alpha)$ (\cref{eq:PR}). 

For $N = 15$ lattice sites, \cref{fig:n15u3.5} and \cref{fig:n15u10} show the comparison between the participation ratio and the two-photon transmission probability for $U = 3.5J$ and $U = 10J$, respectively. It is commonly believed \cite{luitz15,mondragonshem15}, though debatable \cite{deroeck16}, that the presence of interaction between particles causes different eigenstates to localise at different values of $h/J$, forming what is called a mobility edge. This is observed in both quantities at two different interaction strengths in \cref{fig:n15u3.5} and \cref{fig:n15u10}. Furthermore, the two-photon transmission probability $T^{(2)}(\alpha)$ produces very similar behavior as $\log_{d_2}R(\alpha)$. 

Interestingly, when $U = 10J$, 15 eigenstates ($\alpha \geq 106$) localise almost as soon as $h/J>0$. Notably, the two-photon transmission probability $T^{(2)}(\alpha)$ produces the expected behavior of $R$ in this scenario. To understand why this happens, consider the two-particle manifold in Fock state basis: it consists of $N(N-1)/2$ singly occupied (one particle per site) and $N$ doubly occupied (two particles per site) Fock states. In the strongly interacting limit where $U \gg h, J$, the space spanned by the singly occupied states, $\mathcal{S} = \{\ket{i, j}; i<j\}$, is almost disjointed from the space spanned by the doubly occupied states, $\mathcal{D} = \{\ket{i, i}\}$. This causes the eigenstates of the system to be dominated by either the singly occupied or the doubly occupied Fock states, with the latter having higher energies. In this regime, one can decouple $\mathcal{D}$ and $\mathcal{S}$ using the Schrieffer-Wolff transformation to find an effective Hamiltonian in $\mathcal{D}$ which describes the behavior of the $N$ high energy eigenstates (\cref{sec:swt}). It turns out that the effective Hamiltonian in $\mathcal{D}$ to the lowest order resembles that of the AA model with a delocalisation-localisation transition at $h/J = 2J/U$. In contrast, the effective Hamiltonian in $\mathcal{S}$ to the lowest order equals the original Hamiltonian, $H_\text{sys}$. Hence, in the strongly interacting limit, interaction alters the behavior of the high energy eigenstates such that they localised significantly faster as compared to the other eigenstates. Our numerical results in \cref{fig:n15u10}, which has $N =15$ and $U = 10J$, agree perfectly with the theoretical analysis where the 15 high energy eigenstates have a transition at $h/J = 0.2$.
  
\subsection{Scaling with $N$}\label{sec:scaling}
\begin{figure*}[!hbt]
\centering
\sidesubfloat[]{\includegraphics[width=0.3\textwidth]{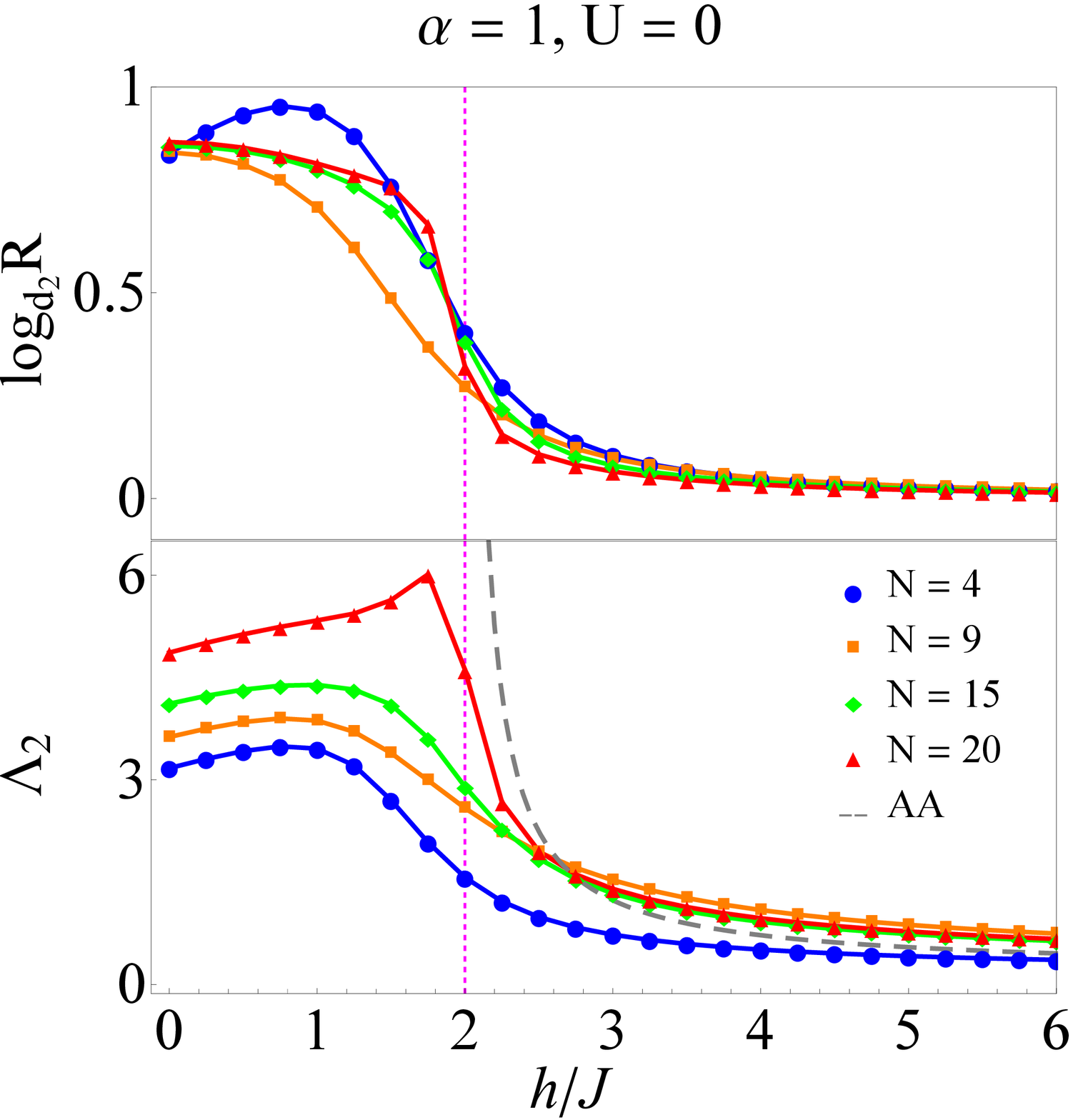}\label{fig:a1u0}}
\sidesubfloat[]{\includegraphics[width=0.3\textwidth]{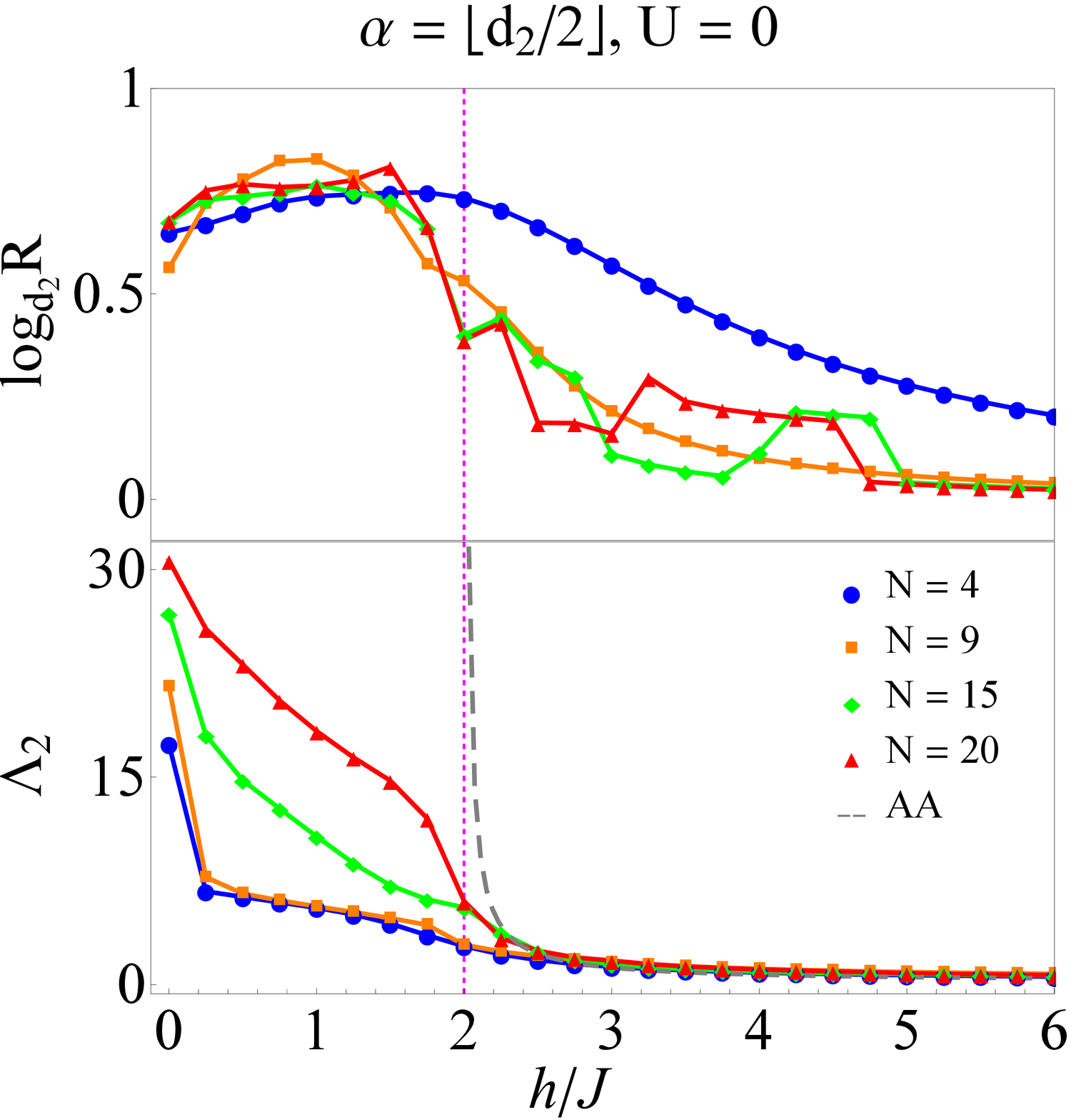}\label{fig:ahd2u0}}
\sidesubfloat[]{\includegraphics[width=0.3\textwidth]{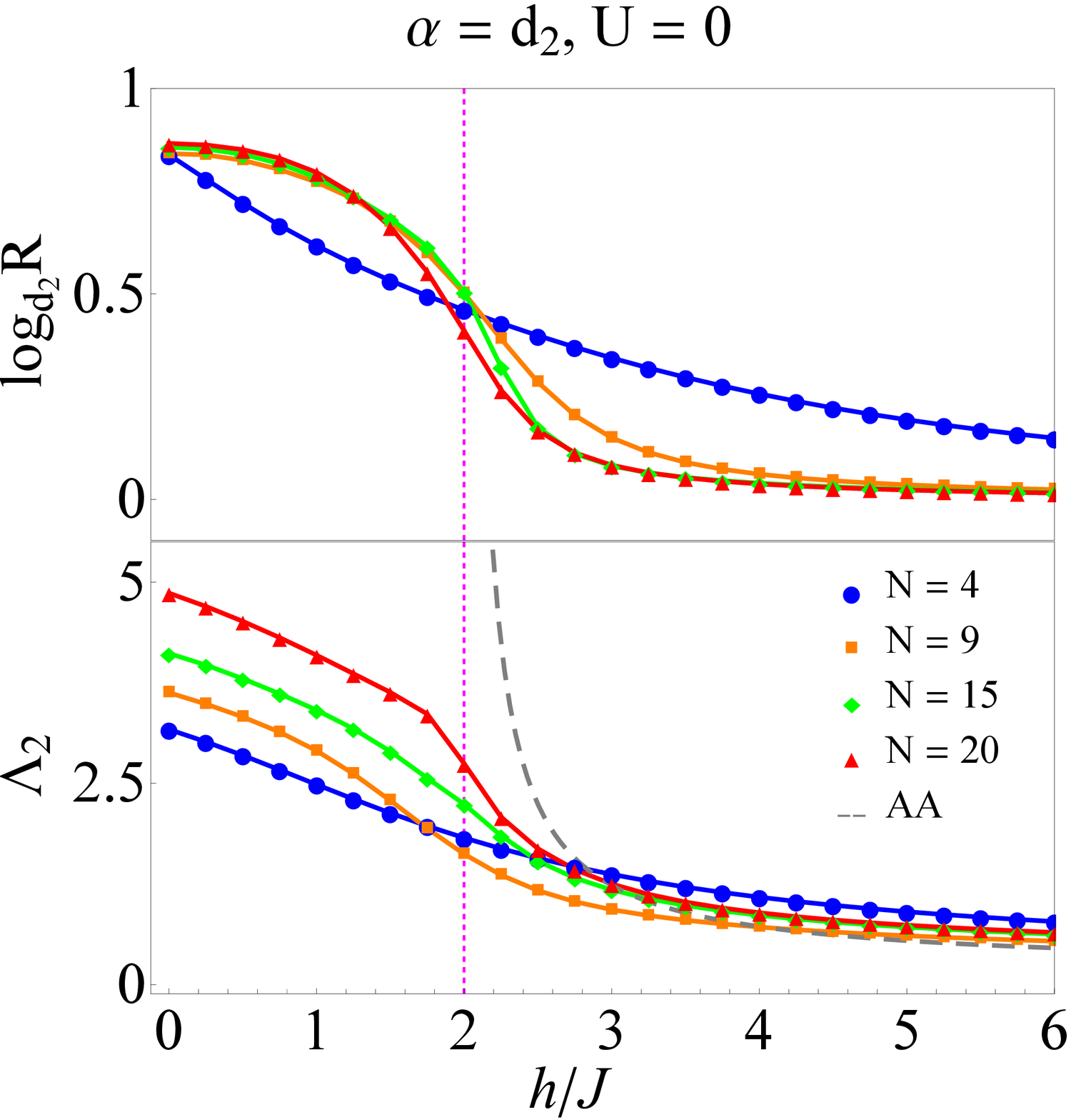}\label{fig:ad2u0}}

\vspace{1em}
\sidesubfloat[]{\includegraphics[width=0.3\textwidth]{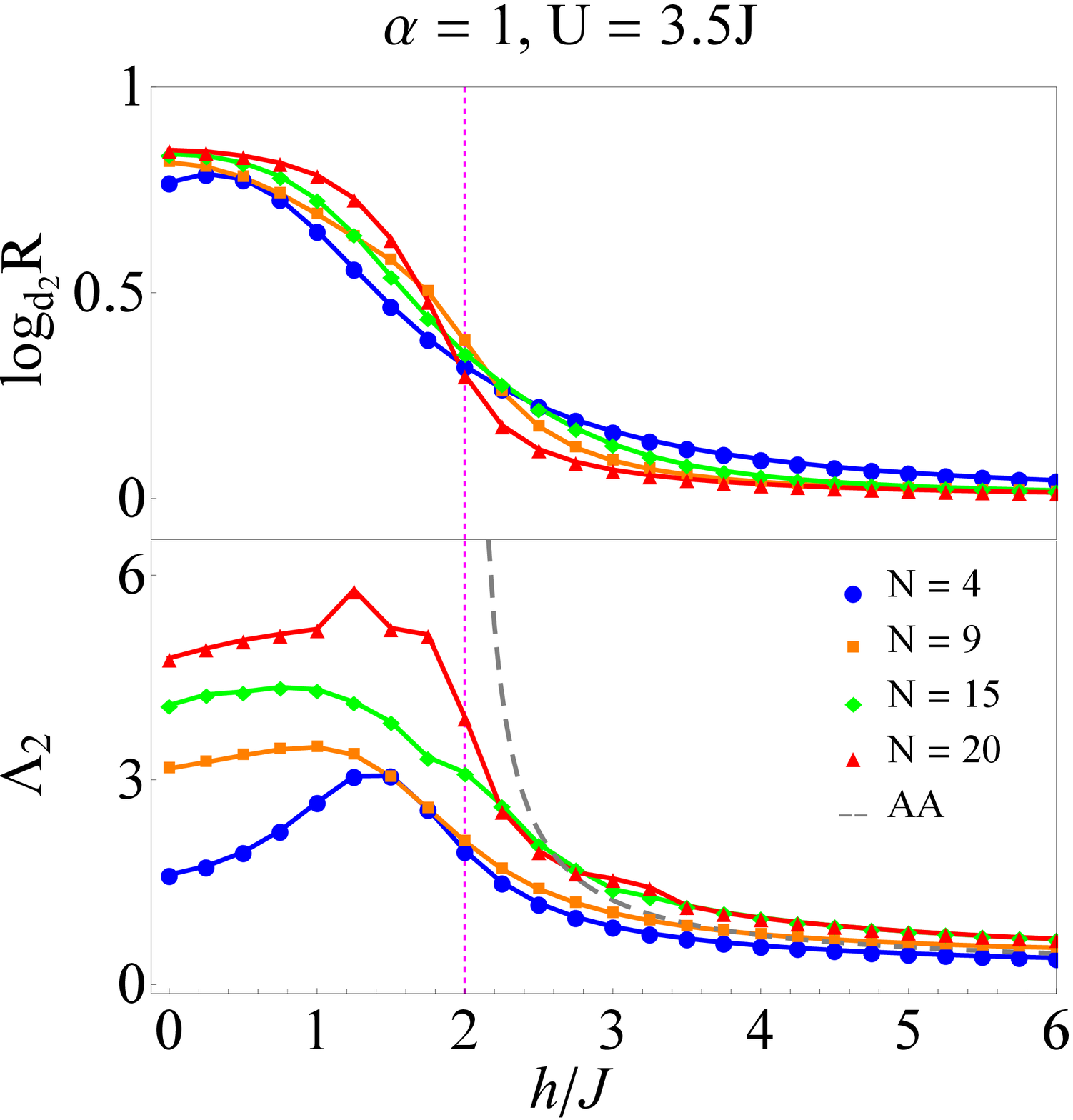}\label{fig:a1u3.5}}
\sidesubfloat[]{\includegraphics[width=0.3\textwidth]{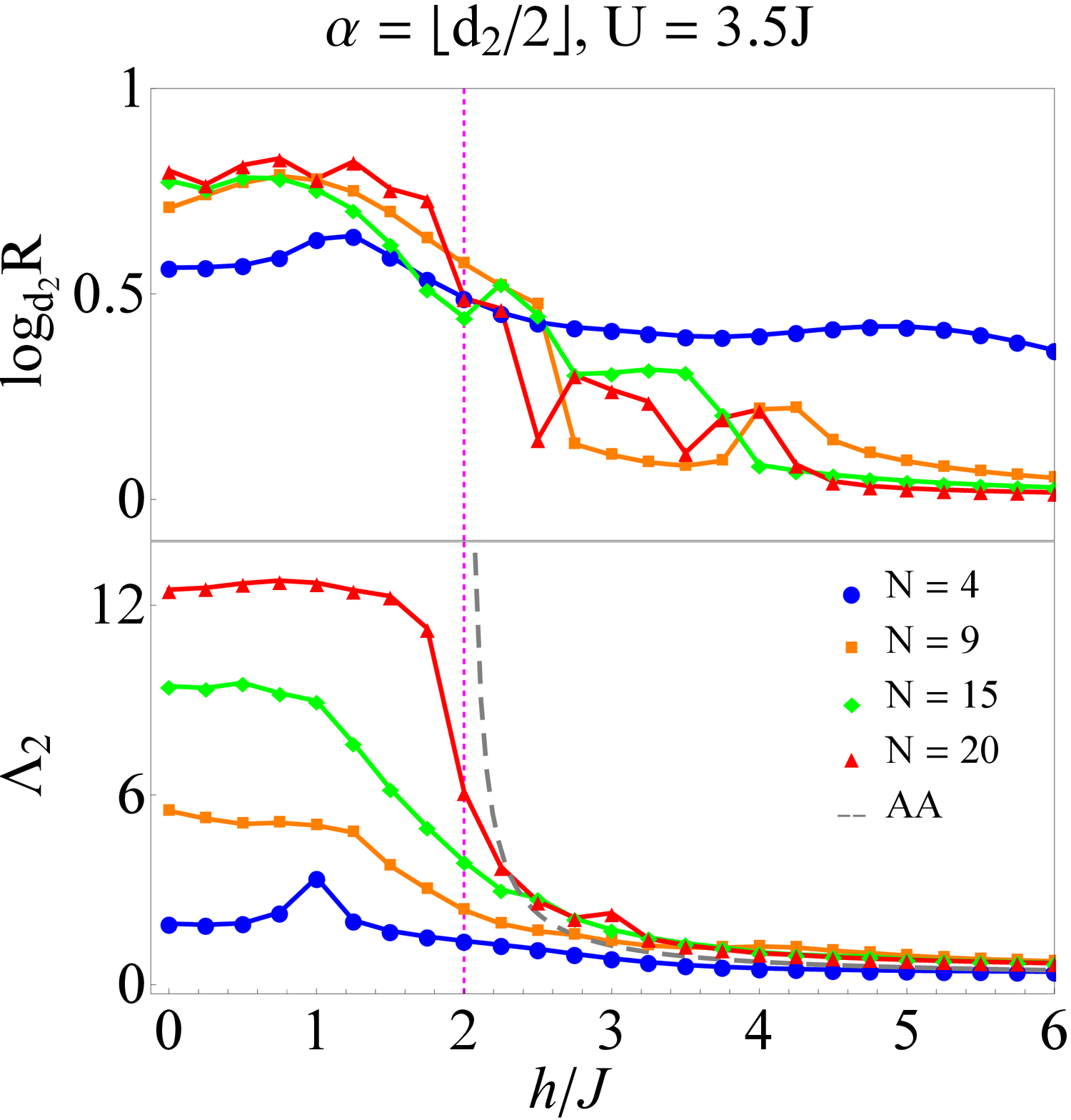}\label{fig:ahd2u3.5}}
\sidesubfloat[]{\includegraphics[width=0.3\textwidth]{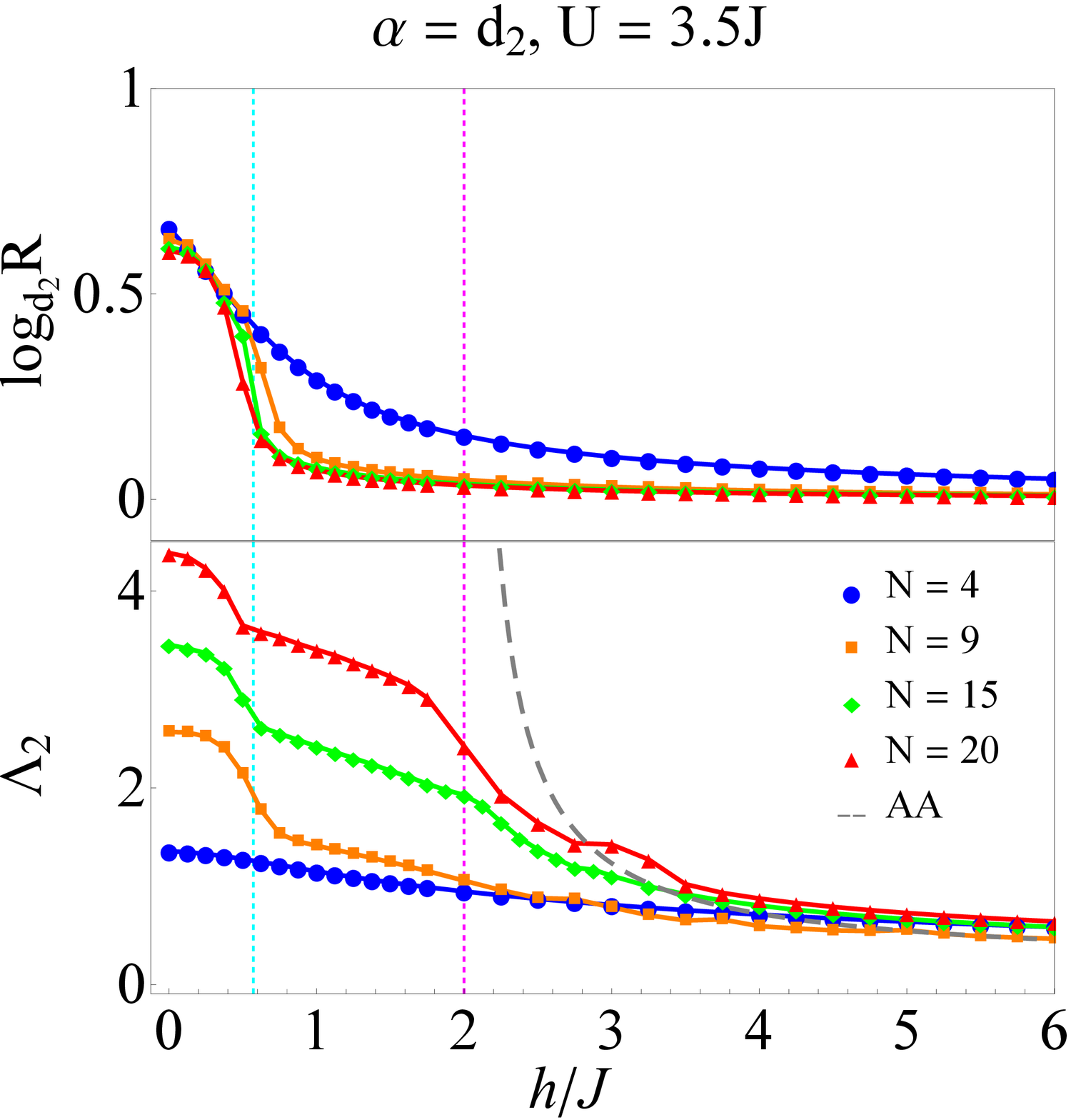}\label{fig:ad2u3.5}}

\vspace{1em}
\sidesubfloat[]{\includegraphics[width=0.3\textwidth]{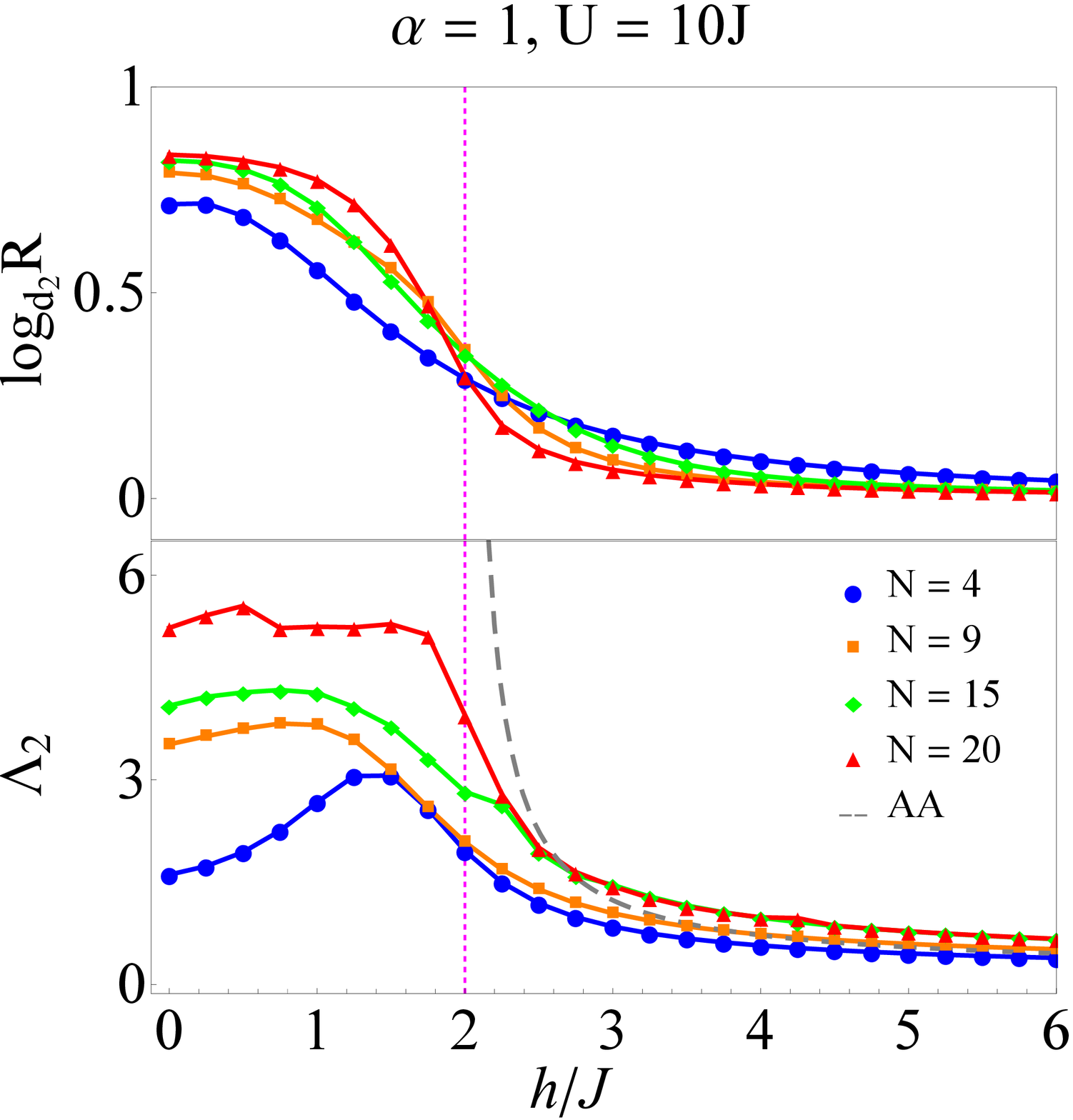}\label{fig:a1u10}}
\sidesubfloat[]{\includegraphics[width=0.3\textwidth]{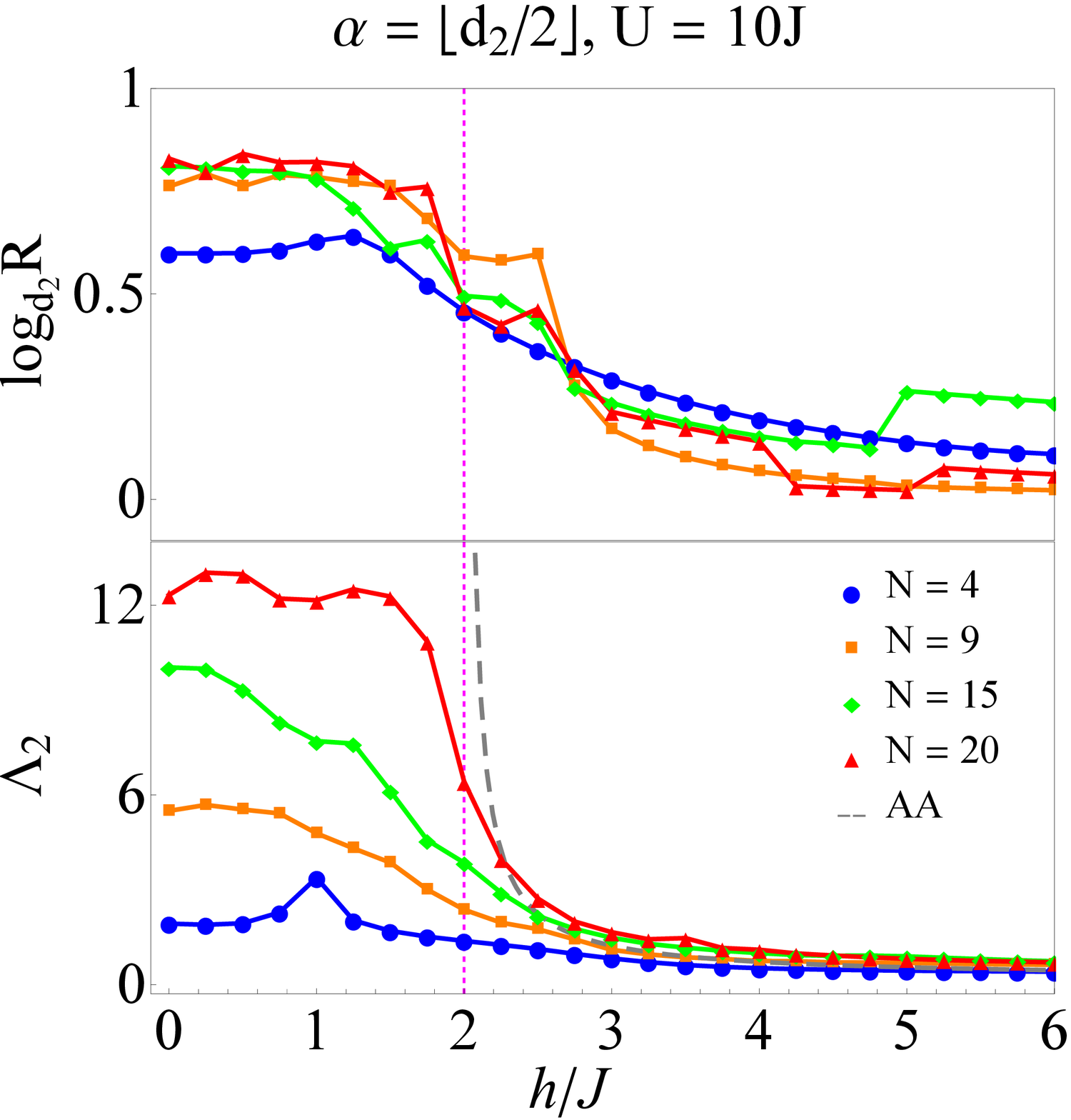}\label{fig:ahd2u10}}
\sidesubfloat[]{\includegraphics[width=0.3\textwidth]{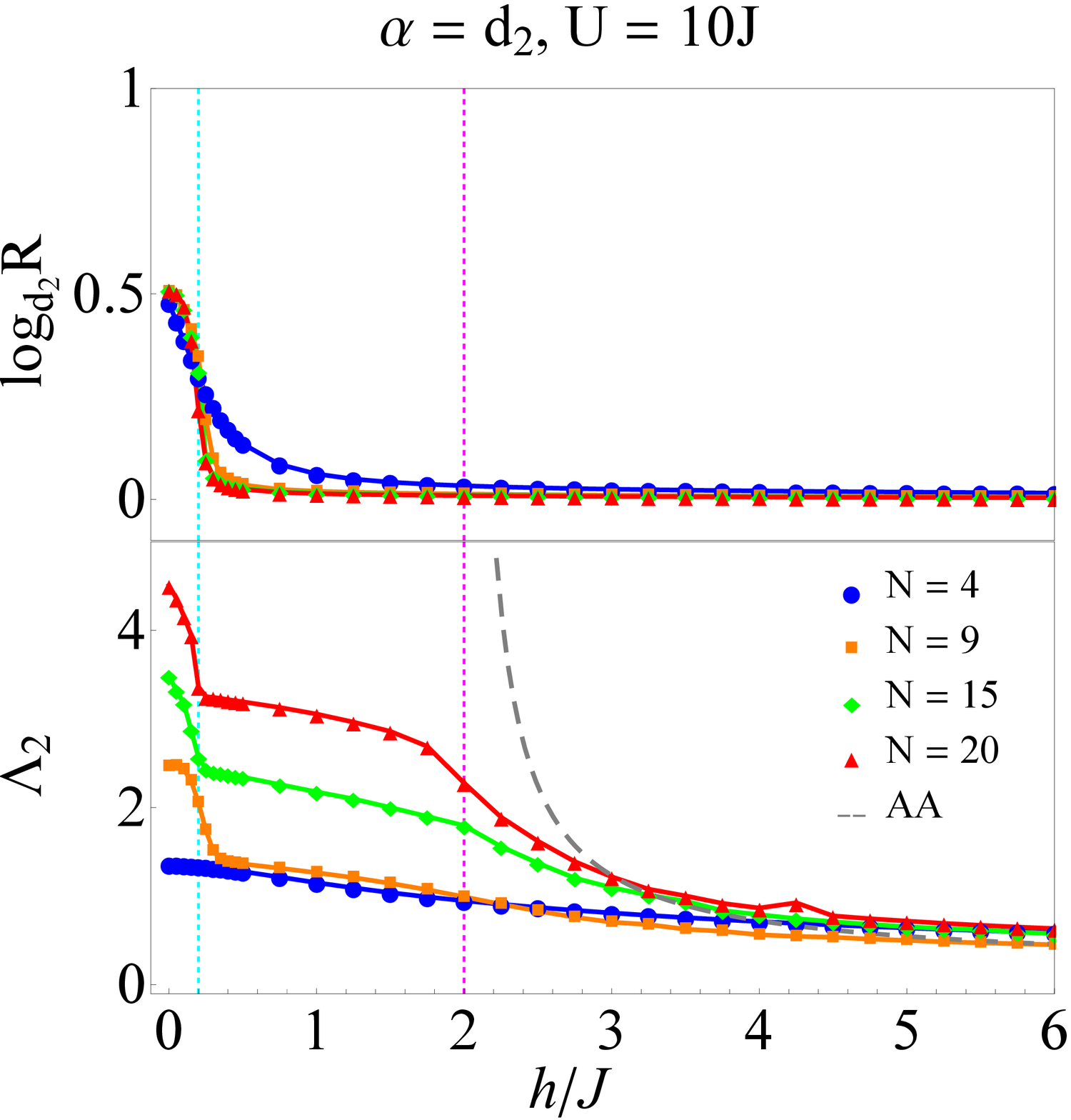}\label{fig:ad2u10}}

\caption{Scaling with $N$: Participation ratio vs effective TIP's localisation length. (a)-(c) $U = 0$ (AA model), (d)-(f) $U = 3.5J$, and (g)-(i) $U = 10J$ for three different eigenstates: $\alpha = 1, \lfloor d_2/2 \rfloor$, and $d_2$. Here, $b = \frac{\sqrt{5} - 1}{2}$ and $\kappa = 0.25J$. Both quantities $\log_{d_2}R$ and $\Lambda_2$, approach the same transition point as $N$ increases. Magenta dashed lines indicate the AA model's predicted transition at $h/J = 2$. For plots of $\Lambda_2$, gray dashed lines labeled AA satisfy the equation $1/\Lambda_2 = 2\ln(h/2J)$ for  $h/J > 2$. It is the infinite $N$ limit of the effective TIP's localisation length in the AA model. For the highest energy eigenstate $\alpha = d_2$, cyan lines indicate the theoretical transition of $h/J = 2J/U$ in the strongly interacting limit of $U >> h,J$. }\label{fig:scaleN}
\end{figure*}

In this section, we study how the participation ratio $R$ and the two-photon transmission probability $T^{(2)}$ scale with the system size, $N$. Following the discussion in previous sections, we will be comparing the quantities, $\log_{d_2}R$ and $\Lambda_2$ (\cref{eq:Lambda2}) instead of the participation ratio and the two-photon transmission probability for ease of analyzing the significance of the results. In the limit of large $N$, $\log_{d_2}R$ approaches the fractal dimensionality of a state. It is equal to 1 when the state is fully delocalised and is equal to 0 when the state is localised \cite{kramer93}. On the other hand, $\Lambda_2$ as a measure of localisation length, should increase with $N$ when a state is delocalised but should approach a small finite value when a state is localised. Furthermore, when $U = 0$ in the AA model, one can show that (\cref{sec:analytic})
\begin{equation}\label{eq:u0Lambda2}
\frac{1}{\Lambda_2}\Big\rvert_{E = E^{(2)}_{\alpha}} \approx \frac{1}{\lambda_1}\Big\rvert_{E = E^{(1)}_{\alpha_1}} + \frac{1}{\lambda_1}\Big\rvert_{E = E^{(1)}_{\alpha_2}},
\end{equation}
where $E_\alpha^{(2)} = E_{\alpha_1}^{(1)} + E_{\alpha_2}^{(1)}$. In the orignal paper by Aubry and Andr{\'e} \cite{aubry80}, they showed that when $h/J > 2$ all states are localised and the single-particle localisation length, $1/\lambda_1 = \ln (h/2J)$. Combining this with \cref{eq:u0Lambda2}, we get that, in the noninteracting regime where $U = 0$,
\begin{equation}\label{eq:linearL2}
\frac{1}{\Lambda_2} \approx 2\ln\left(\frac{h}{2J}\right),
\end{equation} 
for $h/J > 2$.

In \cref{fig:scaleN}, we show how $\log_{d_2}R$ and $\Lambda_2$ scale with the system size $N$ for three different eigenstates $\ket{E^{(2)}_\alpha}$, where $\alpha = 1, \lfloor d_2/2 \rfloor$, and $d_2$ correspond to the lowest, middle, and highest energies, respectively. Magenta dashed lines indicate AA model's predicted transition at $h/J = 2$. For plots of $\Lambda_2$, gray dashed lines labeled AA represent the expected behavior in the AA model as in \cref{eq:linearL2}. For the highest energy eigenstate $\alpha = d_2$, cyan lines indicate the theoretical transition of $h/J = 2J/U$ in the strongly interacting limit of $U >> h,J$.

When $U = 0$, $\log_{d_2} R$ shows a transition for all three eigenstates, with a finite nonzero value when $h/J < 2$ that approaches zero after $h/J = 2$. The plots become more step-like around the transition point, $h/J = 2$, as $N$ increases. On the other hand, $\Lambda_2$ also shows a transition for all three eigenstates with an $N$ dependent large value when $h/J < 2$ which approaches the same small finite value after $h/J = 2$. In the localised phase, $h/J > 2$, the behavior of $\Lambda_2$ approaches the linear case described by \cref{eq:linearL2} as $N$ increases.  

When $U = 3.5J$ and $10J$, we observe that the behavior of the lowest energy eigenstate, $\alpha = 1$, is very similar to that when $U = 0$ for both of the quantities. This is unsurprising as the lowest energy eigenstate is most likely made up of only Fock states in the singly occupied subspace, $\mathcal{S}$. This results in a behavior that is ``linear'', similar to the one when $U = 0$. For $\alpha = \lfloor d_2/2 \rfloor$, the quantity $\log_{d_2} R$ shows a transition around $h/J = 2$ similar to that described for $U = 0$. $\Lambda_2$ in \cref{fig:ahd2u3.5,fig:ahd2u10} displays a more step-like drop around $h/J = 2$ as compared to \cref{fig:ahd2u0}, but both approach the same linear case as the latter when $N$ increases.   

For the highest energy eigenstate, $\alpha = d_2$ when $U \neq 0$, both quantities in \cref{fig:ad2u3.5,fig:ad2u10} show a transition at $h/J = 2J/U$, which is predicted using the Schrieffer-Wolff transformation. This is expected for $U = 10J$, as we have already seen the appearance of the transition when looking at the full density plot in \cref{fig:n15u10}. However, what is also interesting is the reasonably good prediction of the transition even for $U = 3.5J \not\gg J$. If we look at \cref{fig:n15u3.5}, there are a few highly excited eigenstates that localise much faster than the rest at roughly the predicted transition point of $h/J = 2J/U$. The number of such states is smaller than $N$ although the effective Hamiltonian that gives rise to the predicted transition describes the whole doubly occupied subspace $\mathcal{D}$, which has $\dim(\mathcal{D}) = N$. This seems to suggest that the $N$ highest energy eigenstates are not only coming from $\mathcal{D}$, but are also having an increasingly non-negligible contribution from $\mathcal{S}$ as the energy decreases. The assumption that $\mathcal{S}$ and $\mathcal{D}$ are almost disjointed such that all the $N$ highest energy eigenstates have dominating contribution from $\mathcal{D}$ hence breaks down. In spite of that, the effective Hamiltonian for $\mathcal{D}$ from the Schrieffer-Wolff transformation (\cref{sec:swt}) still seems to describe the behavior of high energy eigenstates which have majority contribution from $\mathcal{D}$, just that there are fewer such states when $U > J$ but $U \not\gg J$. 

Further evidence to support the deduction that the $\alpha = d_2$ eigenstate has non-negligible proportion in $\mathcal{S}$ when U = 3.5J (but negligible in $\mathcal{S}$ when $U = 10J$) can be found from the value of $\log_{d_2}R$ when the eigenstate is most delocalised, i.e., when $h = 0$. If an eigenstate sits in $\mathcal{D}$ and is fully delocalised,
\begin{equation}
\log_{d_2}R = \log_{d_2}N \rightarrow \frac{1}{2} \;\text{as}\; N \rightarrow \infty.
\end{equation}
In \cref{fig:ad2u10}, when $U = 10J$, $\log_{d_2}R = 1/2$ when $h = 0$ for all the values of $N$ investigated, whereas in \cref{fig:ad2u3.5} where $U = 3.5J$, $\log_{d_2}R > 1/2$ when $h = 0$ for all values of $N$. Although in \cref{fig:ad2u3.5} $\log_{d_2}R$ decreases as $N$ increases, it is unlikely that it will approach $1/2$ in the limit of infinite $N$ since the value at $N = 15$ almost coincides with that of $N = 20$. This supports the suggestion that there are indeed more than just the Fock states from $\mathcal{D}$ that make up the highest eigenstate, $\alpha = d_2$, when $U = 3.5J$.

Additionally, in the interacting regime, the behavior of $\log_{d_2}R$ is different from $\Lambda_2$ as $h/J$ increases. As shown in \cref{fig:ad2u3.5,fig:ad2u10}, $\log_{d_2}R$ shows a drop from a finite value to zero at $h/J = 2J/U$, whereas $\Lambda_2$ shows a drop at $h/J = 2J/U$ but plateaus in the range $2J/U < h/J < 2$, which eventually falls off to a value independent of $N$ when $h/J > 2$. When $h/J > 2$, $\Lambda_2$ also seems to approach the linear case, given by the gray dashed line, as $N$ increases. From the plots of $\Lambda_2$, it appears that there are two different transitions corresponding to two different mechanisms: one due to interaction at $h/J = 2J/U$, another due to quasiperiodicity at $h/J = 2$. However, this feature is completely absent from $\log_{d_2}R$, where only the transition due to interaction is observed. 

To summarise, the proposed two-photon transmission probability is clearly capable of capturing the delocalisation-localisation transition, and we show this by studying the scaling of its derived quantity, $\Lambda_2$, an effective TIP's localisation length, with the system size $N$. We compare the scaling of $\Lambda_2$ with that of the participation ratio, $\log_{d_2}R$. When $N$ increases, $\log_{d_2}R$ approaches a finite value when the eigenstate is delocalised while dropping more step-like to zero around the transition point, and it remains zero deep in the localised phase. $\Lambda_2$, on the other hand, is $N$ dependent as long as the eigenstate is delocalised but converges to be $N$ independent as the eigenstate transitions from delocalisation to localisation. The convergence of $\Lambda_2$ with $N$ around the transition point also appears to mimic the linear case of the AA model given by \cref{eq:linearL2}. Notably, both quantities capture the theoretically predicted transition point at $h/J = 2J/U$ when the system is in the strongly interacting limit, but $\Lambda_2$ appears to have an intermediate phase at $2J/U < h/J < 2$ which is not observed in $\log_{d_2}R$. The difference in behavior for these two quantities can be understood from the fact that $R$ measures the volume occupied by an eigenstate whereas $\Lambda_2$ characterises the asymptotic decay tail of an eigenstate. Both give a clear change when the system transitions from delocalisation to localisation but they do so by displaying different aspects of the system. 

\subsection{Effect of local losses}\label{sec:loss}
The waveguide QED system that we considered is inherently lossy. To take local losses into account, we couple each site to a harmonic bath. These baths are treated as virtual waveguides that cannot be tracked. If all the sites have the same loss rate $\gamma$, following \cref{sec:smatrix}, the scattering elements can be calculated by considering the effective Hamiltonian
\begin{equation}
\hat{H}_\text{loss} = \hat{H}_\text{eff} - \rmi\frac{\gamma}{2}\sum_{i=1}^N \hat{a}_i^\dagger \hat{a}_i
\end{equation}
with $\hat{H}_\text{eff}$ as defined in \cref{eq:Heff}. The steps for computing the two-photon transmission probability are exactly the same as outlined in \cref{sec:smatrix}, but now with the effective Hamiltonian $\hat{H}_\text{loss}$ instead of $\hat{H}_\text{eff}$.

\begin{figure}[!ht]
\centering
\sidesubfloat[]{
\includegraphics[scale=0.35]{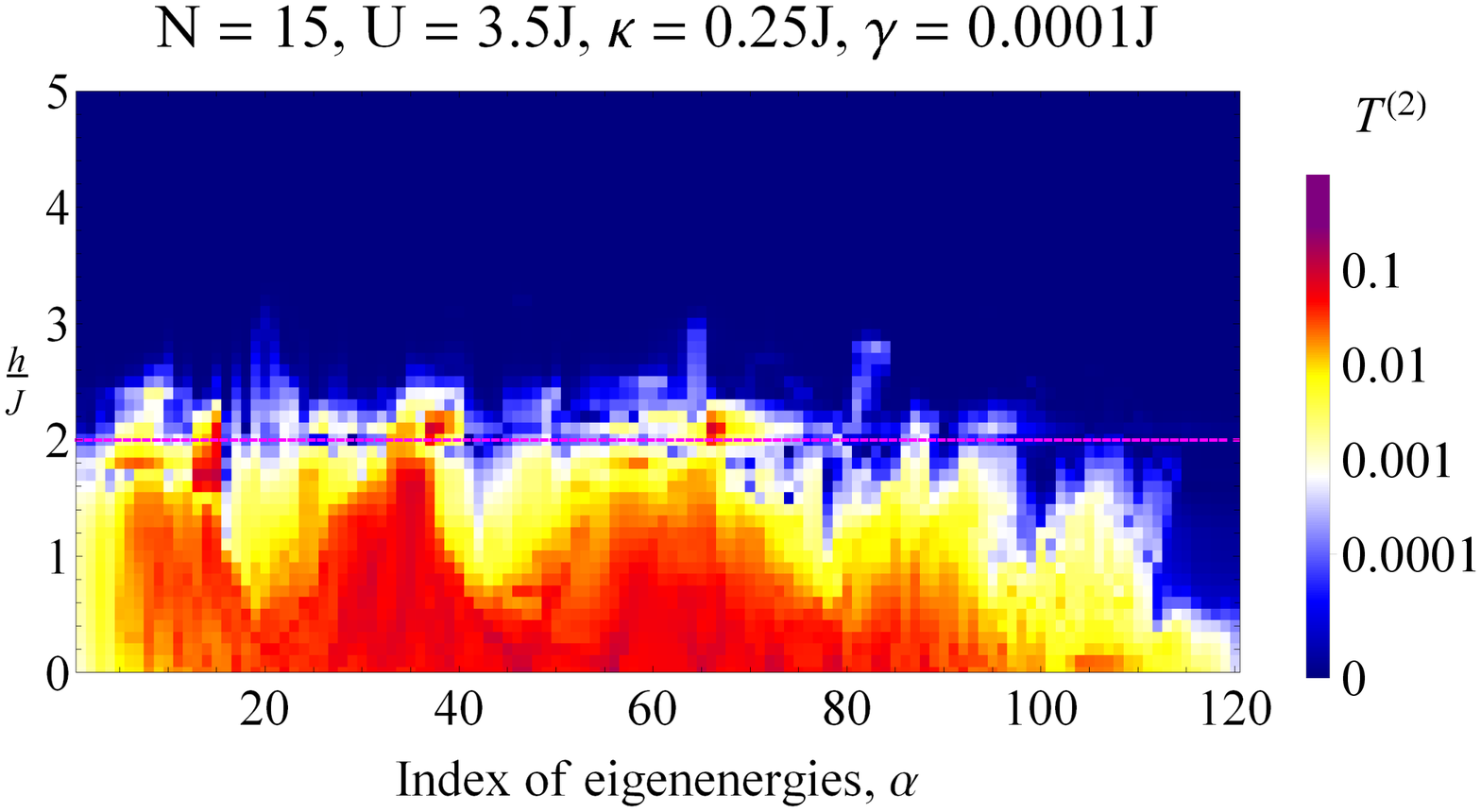}}

\vspace{1em}
\sidesubfloat[]{
\includegraphics[scale=0.35]{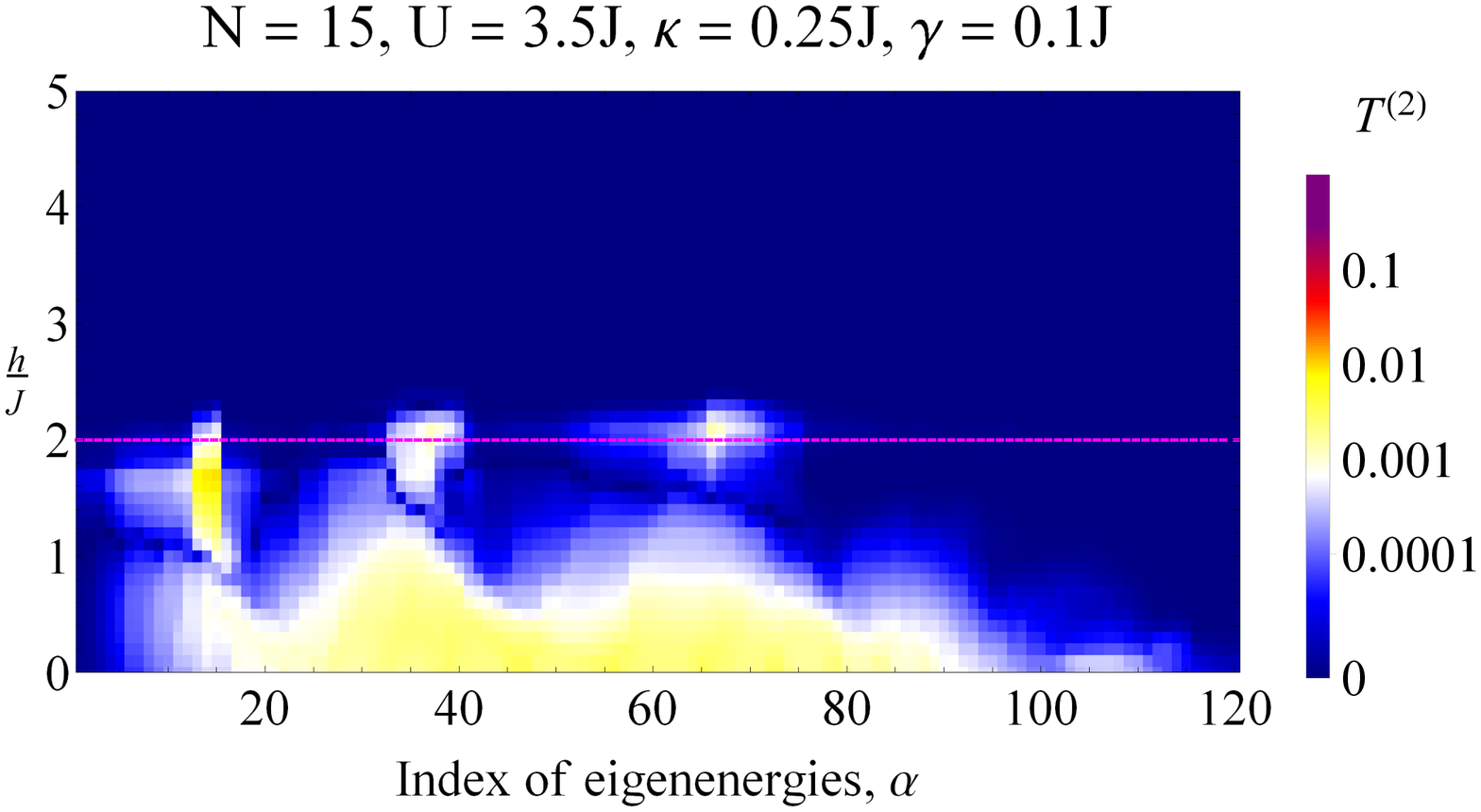}}

\caption{Effect of local losses: Plots of two-photon transmission probability $T^{(2)}(\alpha)$, for two different local loss rates of (a) $\gamma = 0.0001J$ and (b) $\gamma = 0.1J$ for the same waveguide system coupling of $\kappa=0.25J$. Here, $N = 15, U = 3.5J$, and $b = \frac{\sqrt{5} - 1}{2}$. The presence of local losses washes out most of the features observed in the lossless case. However, when the loss rate is much smaller than the coupling strength, $\gamma/\kappa \ll 1$, the mobility edge is still clearly visible.}\label{fig:loss}
\end{figure}

Now, let us consider an interaction strength $U = 3.5J$ and look at the density plots of $T^{(2)}$ in the presence of local losses and consider two cases where $\gamma = 0.0001J$ ($\gamma/\kappa \ll 1$) and $\gamma = 0.1J$ ($\gamma/\kappa \sim 1$) as shown in \cref{fig:loss}. Comparing the lossy case of \cref{fig:loss} with the lossless case of \cref{fig:n15u3.5}, it is clear that the absolute value of $T^{(2)}$ is reduced when the system is lossy. This effect becomes more prominent with increasing loss rate, and when $\gamma/\kappa \sim 1$ previously observed features almost completely disappear, as shown in \cref{fig:loss}. This is unsurprising because the sum of all probabilities of two-photon outputs, including ones through the virtual channels that cannot be tracked, has to be 1. The existence of losses through the virtual channels thus greatly reduces the probability that two photons are transmitted through channel $\mathcal{W}N$, hence washing out the appearance of mobility edge. Although losses break the conservation of photon number, they do not change the intrinsic characteristics of the two-photon transmission from the ideal scenario. This is because if for example, one photon is transmitted and the other is lost, it does not contribute to the measurement of the two-photon transmission probability. Furthermore, the losses we consider are homogeneous, which make these lossy processes homogeneous as well. This results in the reduction of the two-photon transmission probability, but not a change in its characteristics. 

Fortunately, even though the two-photon transmission probability is suppressed by losses, for small loss rate in the strong coupling regime, $\gamma/\kappa \ll 1$, the mobility edge is still visible. The strong coupling regime where $\gamma/\kappa \ll 1$ is within current experimental reach where ratios of $\gamma/\kappa \sim 10^{-4} - 10^{-6}$ have been achieved \cite{aoki06,fink08,hamsen17,moores18}. Similar arguments can also be made even for the general case of an $N$-photon transmission probability.

\subsection{Scattering of coherent light}\label{sec:coherent}
The two-photon transmission probability, $T^{(2)}(\alpha)$ defined in \cref{eq:T2} consists of all scattering processes that are fully resonant. Is the fully resonant condition \cref{eq:resonant} necessary and justified? For instance, if one considers a weak coherent laser field as an input, the two-photon sector of a coherent state produced by a laser corresponds to the scattering process with an input of identical photon momenta. Will the signatures of the underlying localisation transition still be apparent in the transmission spectra or does one have to scatter photons with different frequencies as in the previous section? To answer this, we need to redefine the two-photon transmission probability as follows, with an identical photon momenta input,
\begin{equation}\label{eq:T2c}
T^{(2)}_\text{coh}(\alpha) := P(E_\alpha^{(2)}/2, E_\alpha^{(2)}/2),
\end{equation}
where $P(E_\alpha^{(2)}/2, E_\alpha^{(2)}/2)$ is as defined in \cref{eq:prob} with $k_1 = k_2 = E_\alpha^{(2)}$.

\begin{figure}[!h]
\centering
\includegraphics[scale=0.35]{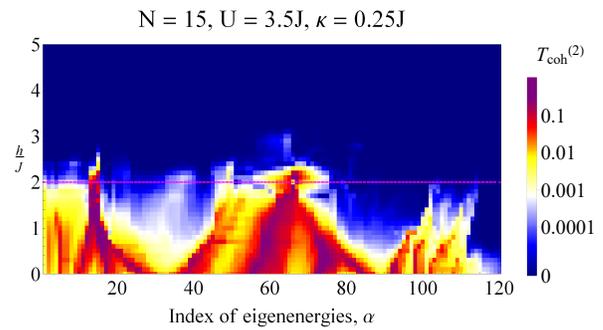}

\caption{Coherent state input: Plot of the two-photon transmission probability with identical photon momenta input (a small coherent state), $T^{(2)}_\text{coh}(\alpha)$. Here, $N = 15, U = 3.5J, b = \frac{\sqrt{5} - 1}{2}$, and $\kappa = 0.25J$. $T^{(2)}_\text{coh}$ behaves very differently from $T^{(2)}$ and $\log_{d_2}R$.}\label{fig:T2c}
\end{figure}

\Cref{fig:T2c} shows the density plot of $T^{(2)}_\text{coh}$ when $N = 15$ and $U = 3.5J$. By comparing $T^{(2)}_\text{coh}$ and $T^{(2)}$ from \cref{fig:T2c} and \cref{fig:n15u3.5} respectively, it can be seen that the coherent-state-like input that does not fulfill the resonant condition, gives rise to a qualitatively different behavior. It can be seen that the density plot of $T^{(2)}_\text{coh}$ does not resemble that of $R$. The choice of input state therefore affects the transmission probability greatly. The numerical evidence presented here together with the quantitative argument in \cref{sec:relation} support the need to impose the resonant condition \cref{eq:resonant}. Advantages of the fully resonant scenario over the coherent input scenario were also previously investigated in the Bose-Hubbard dimer \cite{lee15}. 

\section{Conclusion}
We analysed how correlated photon transport in a nonlinear photonic lattice with disorder can be used to probe signatures of localisation transition. By merging scattering theory with input output formalism, we calculated the transmission spectra for various values of interaction and disorder strength and found patterns very similar to the well known participation ratio of the eigenstates characterizing the delocalisation-localisation transition. Interestingly, the two-photon transmission probability shows how two competing mechanisms, disorder and interaction, influence the transition in the highly energetic eigenstates, which is not seen by just observing the participation ratio. We also discussed how, in experimental situations where the local emitters might exhibit additional local losses, the two-photon transmission still performs well as long as the loss rate is much smaller than the waveguide-system coupling rate. As future work, it would be interesting to study the scaling of the method for larger numbers of photons, aiming at resolving more eigenstates and probing other exotic phenomena. For that, the development of an approximate method to deal with the cumbersome nature of the scattering approach will be needed.

\section{Acknowledgements}
This research is supported by the Singapore Ministry of Education Academic Research Fund Tier 3 (Grant No. MOE2012-T3-1-009), the National Research Foundation, Prime Minister’s Office, Singapore, and the Ministry of Education, Singapore, under the Research Centres of Excellence programme. This research was also partially funded by the Polisimulator project cofinanced by Greece and the EU Regional Development Fund. V. M. B. acknowledges fruitful discussions with W. J. Munro. T. F. S. acknowledges fruitful discussions with H. C. J. Gan.

\appendix
\section{Quantitative details of the relationship between transmission probability and participation ratio}\label{sec:analytic}
If we approximate the two-photon input as delta pulses, i.e., $\chi_k(q) = \delta(k - q)$, the two-photon transmission probability \cref{eq:prob} for a given input $\ket{\text{in}; k_1, k_2}$ can be expressed as
\begin{align}\label{eq:probdelta}
&P(k_1, k_2) \notag\\
= &|G(k_1; k_1)|^2|G(k_2; k_2)|^2 \notag \\
&+ 2 \text{Re}\left[G^*(k_1; k_1)G^*(k_2; k_2)G(k_1, k_2; k_1, k_2)\right] \notag\\
&+ \frac{1}{2}\int \rmd p |G(p, k_1 + k_2 - p; k_1, k_2)|^2,
\end{align}
where we have used $G(p_1; k_1)$ and $G(p_1, p_2; k_1, k_2)$ to represent the two-point and four-point Green's functions defined in \cref{eq:2point} and \cref{eq:4point} without the delta function factor in them. 

In the linear case when $U = 0$, the four-point Green's function, $G(p_1, p_2; k_1, k_2)$ vanishes, as it describes only nonlinear effects. Hence, \cref{eq:probdelta} will become $P(k_1, k_2) = |G(k_1; k_1)|^2|G(k_2; k_2)|^2$. The two-photon transmission probability of eigenstate $\ket{E_\alpha^{(2)}}$, $T^{(2)}(\alpha)$, is then given by
\begin{align}\label{eq:t2aU0}
&T^{(2)}(\alpha) \notag \\
= &\frac{1}{d_1}\sum_{\mu = 1}^{d_1} P(E_\mu^{(1)}, E_\alpha^{(2)} - E_\mu^{(1)}) \notag\\
= &\frac{1}{d_1}\sum_{\mu = 1}^{d_1} \big|G( E_\mu^{(1)}; E_\mu^{(1)})G(E_\alpha^{(2)} - E_\mu^{(1)}; E_\alpha^{(2)} - E_\mu^{(1)})\big|^2 \notag\\
\approx &\frac{1}{2\pi}\int \rmd k \big|G( E_\mu^{(1)}; E_\mu^{(1)})G(E_\alpha^{(2)} - E_\mu^{(1)}; E_\alpha^{(2)} - E_\mu^{(1)})\big|^2 \notag\\
\approx &\kappa^4 \sum_{\mu, \nu, \mu', \nu'}  \left(\frac{i}{\xi_{\nu'}^{(1)*} - \xi_{\nu}^{(1)}} + \frac{i}{\xi_{\mu'}^{(1)*} - \xi_{\mu}^{(1)}}\right) \notag \\
&\hspace{1.2em} \times\left(\frac{\braket{0|\hat{a}_{N}|\xi^{(1)}_{\nu}}\braket{\bar{\xi}^{(1)}_{\nu}|\hat{a}_{1}^\dagger|0}\braket{0|\hat{a}_{N}|\xi^{(1)}_{\mu}}\braket{\bar{\xi}^{(1)}_{\mu}|\hat{a}^\dagger_{1}|0}}{E_\alpha^{(2)} - \xi^{(1)}_{\mu} - \xi^{(1)}_{\nu}}\right) \notag\\
&\hspace{1.2em} \times\left(\frac{\braket{0|\hat{a}_{N}|\xi^{(1)}_{\nu'}}^*\braket{\bar{\xi}^{(1)}_{\nu'}|\hat{a}^\dagger_{1}|0}^*\braket{0|\hat{a}_{N}|\xi^{(1)}_{\mu'}}^*\braket{\bar{\xi}^{(1)}_{\mu'}|\hat{a}^\dagger_{1}|0}^*}{E_\alpha^{(2)} - \xi^{(1)*}_{\mu'} - \xi^{(1)*}_{\nu'}}\right),
\end{align}
where we have approximated the summation in the definition of $T^{(2)}$ by an integral.

When $U \neq 0$, the four-point Green's function is nonzero. However, due to the resonant condition in \cref{eq:resonant} where only fully resonant two-photon transitions are considered, we only have to evaluate the expression 
\begin{align}\label{eq:papprox}
&P(E_\mu^{(1)}, E_\alpha^{(2)} - E_\mu^{(1)}) \notag \\
\approx &\frac{1}{2}\int \rmd p |G(p, E_\alpha^{(2)} - p; E_\mu^{(1)}, E_\alpha^{(2)} - E_\mu^{(1)})|^2 \notag \\
\approx &\frac{1}{2}\int \rmd p |G^{(2)}(p, E_\alpha^{(2)} - p; E_\mu^{(1)}, E_\alpha^{(2)} - E_\mu^{(1)})|^2.
\end{align}

The first approximation comes from the fact that the amplitudes of the first two terms in \cref{eq:probdelta} are much smaller compared to the third term, since the former contain a factor $G(E_\alpha^{(2)} - E_\mu^{(1)}; E_\alpha^{(2)} - E_\mu^{(1)})$ which corresponds to one-photon transition that is off resonant as long as $U \neq 0$. Next, notice that the four-point Green's function is a sum of two diagrams: $G^{(1)}$ which describes two one-photon transitions and $G^{(2)}$ which describes a single two-photon transition (\cref{fig:scattd4}). By the same argument that the off resonant one-photon transition factor in $G^{(1)}$ will make its amplitude much smaller compared to the fully resonant $G^{(2)}$, we make the approximation in the last line of \cref{eq:papprox}. 

By using \eqref{eq:papprox} and approximating the summation by an integral, we have
\begin{widetext}
\begin{align}\label{eq:t2aUneq0}
&T^{(2)}(\alpha) \notag \\
= &\frac{1}{d_1}\sum_{\mu = 1}^{d_1} P(E_\mu^{(1)}, E_\alpha^{(2)} - E_\mu^{(1)}) \notag \\
\approx &\frac{1}{2d_1}\sum_{\mu = 1}^{d_1}\int \rmd p |G^{(2)}(p, E_\alpha^{(2)} - p; E_\mu^{(1)}, E_\alpha^{(2)} - E_\mu^{(1)})|^2 \notag \\
\approx &\frac{1}{4\pi}\int \rmd k \int \rmd p |G^{(2)}(p, E_\alpha^{(2)} - p; k, E_\alpha^{(2)} - k)|^2 \notag \\
\approx &\frac{\kappa^4}{\pi} \sum_{\mu, \nu, \beta, \mu', \nu', \beta'} \left(\frac{\braket{0|\hat{a}_{N}|\xi^{(1)}_{\nu}}\braket{\bar{\xi}^{(1)}_{\nu}|\hat{a}_{N}|\xi^{(2)}_{\beta}}\braket{\bar{\xi}^{(2)}_{\beta}|\hat{a}^\dagger_{1}|\xi^{(1)}_{\mu}}\braket{\bar{\xi}^{(1)}_{\mu}|\hat{a}^\dagger_{1}|0}}{E_\alpha^{(2)} - \xi^{(2)}_{\beta}}\right)\frac{i}{\xi_{\nu'}^{(1)*} - \xi_{\nu}^{(1)}} \notag \\
&\hspace{7.3em} \times\left(\frac{\braket{0|\hat{a}_{N}|\xi^{(1)}_{\nu'}}^*\braket{\bar{\xi}^{(1)}_{\nu'}|\hat{a}_{N}|\xi^{(2)}_{\beta'}}^*\braket{\bar{\xi}^{(2)}_{\beta'}|\hat{a}^\dagger_{1}|\xi^{(1)}_{\mu'}}^*\braket{\bar{\xi}^{(1)}_{\mu'}|\hat{a}^\dagger_{1}|0}^*}{E_\alpha^{(2)} - \xi^{(2)*}_{\beta'}}\right)\frac{i}{\xi_{\mu'}^{(1)*} - \xi_{\mu}^{(1)}}.
\end{align}
\end{widetext}

Next, we write the matrix element of the resolvent operator, $\hat{\mathcal{G}}(E) := \left(E - \hat{H}_\text{sys}\right)^{-1}$ in \cref{eq:ll2} in spectral representation as
\begin{align}\label{eq:ResNN11}
\braket{N, N|\hat{\mathcal{G}}(E)|1, 1} &= \braket{0|\frac{a_N^2}{\sqrt{2}}\left(E - \hat{H}_\text{sys}\right)^{-1}\frac{a_1^{\dagger 2}}{\sqrt{2}}|0} \notag \\
&= \sum_{\beta} \frac{\braket{0|\hat{a}_N^2|E_\beta^{(2)}}\braket{E_\beta^{(2)}|\hat{a}_1^{\dagger 2}|0}}{2\left(E - E_{\beta}^{(2)}\right)}.
\end{align}
When $U = 0$, it can be expressed in terms of single-particle eigenenergies and eigenstates as
\begin{align}\label{eq:ResNN11U0}
&\braket{N, N|\hat{\mathcal{G}}(E)|1, 1} \notag \\
= &\sum_{\beta_1, \beta_2} \frac{\braket{0|\hat{a}_N|E_{\beta_1}^{(1)}}\braket{E_{\beta_1}^{(1)}|\hat{a}_1^\dagger|0}\braket{0|\hat{a}_N|E_{\beta_2}^{(1)}}\braket{E_{\beta_2}^{(1)}|\hat{a}_1^\dagger|0}}{E - E^{(1)}_{\beta_1} - E^{(1)}_{\beta_2}}.
\end{align}
Comparing \cref{eq:t2aU0} with \cref{eq:ResNN11U0} and \cref{eq:t2aUneq0} with \cref{eq:ResNN11}, we see that the transmission probability is roughly proportional to the modulus squared of the matrix element of the resolvent operator as given by \cref{eq:ReseqT2}, i.e., $ T^{(2)} \approx \mathcal{C} |\braket{N, N|\hat{\mathcal{G}}|1, 1}|^2$. 

Similarly, the relationship between the participation ratio $R$ and the resolvent operator $\hat{\mathcal{G}}$ can been seen by expressing the diagonal of $\hat{\mathcal{G}}$ in spectral representation, 
\begin{align}\label{eq:Resdiag}
\braket{i, j|\hat{\mathcal{G}}(E)|i, j} &= \braket{0|a_i a_j\left(E - \hat{H}_\text{sys}\right)^{-1}a_i^\dagger a_j^\dagger|0} \notag \\
&= \sum_{\beta} \frac{\braket{i, j|E_\beta^{(2)}}\braket{E_\beta^{(2)}|i, j}}{E - E_{\beta}^{(2)}} \notag \\
&= \sum_{\beta} \frac{|c_{i, j}^\beta|^2}{E - E_{\beta}^{(2)}}.
\end{align}
Finally, comparing \cref{eq:PR} with \cref{eq:Resdiag}, we deduce the relation \cref{eq:ReseqPR-1}, which is $R^{-1}(\alpha) = \sum_{i \leq j}\left|\braket{i, j|\hat{\mathcal{G}}(E^{(2)}_\alpha)|i, j}\right|^2$.

Another natural question one would ask when looking at \cref{eq:ResNN11U0,eq:t2aU0} in the linear case of $U = 0$ is how similar they are with the matrix element in the definition of the single-particle localisation length $\lambda_1$ in \cref{eq:ll1}. From \cref{eq:ResNN11U0}, it is obvious that 
\begin{equation}
\braket{N, N|\hat{\mathcal{G}}(E^{(2)}_\alpha)|1, 1} = \braket{N|\hat{\mathcal{G}}(E^{(1)}_{\alpha_1})|1}\braket{N|\hat{\mathcal{G}}(E^{(1)}_{\alpha_2})|1},
\end{equation}
where $E^{(2)}_\alpha = E^{(1)}_{\alpha_1} + E^{(1)}_{\alpha_2}$. Hence, 
\begin{equation}\label{eq:t2ll1rel}
T^{(2)}(\alpha) \approx \mathcal{C} |\braket{N|\hat{\mathcal{G}}(E^{(1)}_{\alpha_1})|1}\braket{N|\hat{\mathcal{G}}(E^{(1)}_{\alpha_2})|1}|^2,
\end{equation}
where $\mathcal{C}$ is a constant. Combining \cref{eq:t2ll1rel} with the definition of the effective TIP's localisation length, \cref{eq:Lambda2}, and the definition of the single-particle localisation length, \cref{eq:ll1}, one will recover the relationship
\begin{equation}
\frac{1}{\Lambda_2}\Big\rvert_{E = E^{(2)}_{\alpha}} \approx \frac{1}{\lambda_1}\Big\rvert_{E = E^{(1)}_{\alpha_1}} + \frac{1}{\lambda_1}\Big\rvert_{E = E^{(1)}_{\alpha_2}},
\end{equation}
as in \cref{eq:u0Lambda2}. 

\section{Schrieffer-Wolff transformation}\label{sec:swt}
The two-particle manifold of $\hat{H}_\text{sys}$ in \cref{eq:AAwint} is made up of singly occupied Fock states, $\mathcal{S} = \{\ket{i, j}; i<j\}$, and doubly occupied Fock states, $\mathcal{D} = \{\ket{i, i}\}$. In the strongly interacting regime where $U \gg J, h$, the two subspaces, $\mathcal{S}$ and $\mathcal{D}$, are almost disjointed from each other. In this regime, the hopping term can be treated as a small perturbation, which allows us to apply the Schrieffer-Wolff transformation \cite{schrieffer66}. Under the unitary transformation, we can decouple the two subspaces in the resultant Hamiltonian, $\hat{H}' = \rme^{\rmi \hat{S}} \hat{H}_\text{sys} \rme^{-\rmi \hat{S}}$, where $\hat{S} = \hat{S}^\dagger$ is a Hermitian operator. Our final aim is to obtain the effective hopping (to the lowest order) within the doubly occupied Fock states subspace, $\mathcal{D}$.

We write $\hat{H}_\text{sys} = \hat{H}_0 + \lambda \hat{V}$ with 
\begin{equation}
\hat{H}_0 = \sum_{j = 1}^N \left(\epsilon_j \hat{a}_j^\dagger \hat{a}_j + \frac{U}{2}\hat{a}_j^\dagger \hat{a}_j^\dagger \hat{a}_j \hat{a}_j\right),
\end{equation}
which has Fock states $\ket{i, j}$ as eigenstates with eigenenergies $E^0_{ij} = \epsilon_i + \epsilon_j + U \delta_{ij}$ and 
\begin{equation}
\lambda\hat{V} = J \sum_{j = 1}^{N - 1} \left(\hat{a}_j^\dagger \hat{a}_{j + 1} + \text{H.c.}\right),
\end{equation}
which is the small perturbation.

Using Baker-Campbell-Hausdorff formula for the resultant Hamiltonian,
\begin{equation}
\hat{H}' = \hat{H} + \big[\rmi \hat{S}, \hat{H}\big] + \frac{1}{2}\big[\rmi \hat{S}, \big[\rmi \hat{S}, \hat{H}\big]\big] + \cdots,
\end{equation}
and writing $\hat{S}$ in increasing order of $\lambda$,
\begin{equation}
\hat{S} = \lambda\hat{S}_1 + \lambda^2\hat{S}_2 + \cdots,
\end{equation}
the resultant Hamiltonian in increasing order of $\lambda$ is given by
\begin{align}
\hat{H}'_0 &= \hat{H}_0 \\
\hat{H}'_1 &= \lambda\hat{V} + \big[\rmi\lambda\hat{S}_1, \hat{H}_0\big] \label{eq:s1}\\
\hat{H}'_2 &= \big[\rmi\lambda\hat{S}_1, \lambda\hat{V}\big] + \big[\rmi\lambda^2\hat{S}_2, \hat{H}_0\big] + \frac{1}{2}\big[\rmi\lambda\hat{S}_1, \big[\rmi\lambda\hat{S}_1, \hat{H}_0\big]\big] \label{eq:s2}\\
&\shortvdotswithin{=} \notag
\end{align}
where $\hat{H}'_n$ is of order $\lambda^n$ and the zeroth order of $\hat{S}$ is set to zero so that $\hat{H}' = \hat{H}$ at zeroth order. 

In order to decouple $\mathcal{S}$ and $\mathcal{D}$, we require that 
\begin{align}\label{eq:decouple}
\braket{l, m|\hat{H}'|i, i} &= 0,  &\forall \ket{l, m} \in \mathcal{S},\; \forall \ket{i, i} \in \mathcal{D}.
\end{align}

Combining \cref{eq:decouple} with the first-order equation \cref{eq:s1}, we have
\begin{equation}
\braket{l, m|\rmi\lambda\hat{S}_1|i, i} = \frac{\braket{l, m|\lambda\hat{V}|i, i}}{E^0_{lm} - E^0_{ii}},
\end{equation}
$\forall \ket{l, m} \in \mathcal{S}$ and $\forall \ket{i, i} \in \mathcal{D}$. Besides this constraint, there are no restrictions on the other matrix elements, and hence without any loss of generality we can set the rest to zero. 

Since the only relevant nonzero matrix elements of $\lambda\hat{V}$ are
\begin{equation}
\braket{i, i+1|\lambda\hat{V}|i+1, i+1} = \braket{i, i+1|\lambda\hat{V}|i, i} = \sqrt{2}J,
\end{equation}
for $i = 1, \dots, N-1$, the first order term of $\hat{S}$ is
\begin{multline}
\rmi\lambda\hat{S}_1 = \sqrt{2}J\sum_{j = 1}^{N - 1} \Bigg(\frac{\ket{i+1, i+1}\bra{i, i+1}}{\epsilon_{i+1} - \epsilon_i + U} \\+ \frac{\ket{i, i}\bra{i, i+1}}{\epsilon_i - \epsilon_{i+1} + U}  - \text{H.c.}\Bigg).
\end{multline}

Similarly, we can use \cref{eq:decouple} together with the second-order equation \cref{eq:s2} to find the constraint for $\hat{S}_2$ such that $\mathcal{S}$ and $\mathcal{D}$ are decoupled. Since the constraint will only involve cross terms from those two subspaces, the other matrix elements can be set to zero as before. 

As we are mainly interested in obtaining the effective hopping term within $\mathcal{D}$, the relevant matrix element to consider is $\braket{i, i|\hat{H}'|i+1, i+1}$. To the lowest order, it is given by
\begin{align}
&\braket{i, i|\hat{H}'_2|i+1, i+1} \notag\\
= &\braket{i, i|\big[\rmi\lambda\hat{S}_1, \lambda\hat{V}\big]|i+1, i+1} \notag\\
&+ \frac{1}{2}\braket{i, i|\big[\rmi\lambda\hat{S}_1, \big[\rmi\lambda\hat{S}_1, \hat{H}_0\big]\big]|i+1, i+1} \notag\\
= &\frac{2J^2U}{U^2 - (\epsilon_{i+1} - \epsilon_i)^2},
\end{align}
where $\braket{i, i|\big[\rmi\lambda^2\hat{S}_2, \hat{H}_0\big]|i+1, i+1} = 0$ by the choice of $\hat{S}_2$. Therefore, to the lowest order, the effective Hamiltonian for the subspace $\mathcal{D}$ is 
\begin{multline}
\hat{H}'_{\mathcal{D}} = \sum_{j = 1}^N \left(h_\mathcal{D}\cos(2\pi bj) + U\right)\ket{j, j}\bra{j, j} \\+ \sum_{j = 1}^{N - 1} J_{\mathcal{D}j}\left(\ket{j, j}\bra{j+1, j+1} + \text{H.c.}\right).
\end{multline}
where the effective quasiperiodicity strength $h_\mathcal{D} = 2h$ and effective hopping strength $J_{\mathcal{D}j} = 2J^2U/(U^2 - (\epsilon_{j+1} - \epsilon_j)^2)$. 

Since $U \gg h$, the effective hopping strength within $\mathcal{D}$ is approximately constant, i.e., $J_{\mathcal{D}j} = J_{\mathcal{D}} := 2J^2/U \;\forall j$. Under this condition, $\hat{H}'_{\mathcal{D}}$ resembles the AA model with a constant shift of $U$ on its onsite energy which can be ignored. Applying the same argument as in the AA model, the delocalisation-localisation transition should happen at $h_\mathcal{D}/J_\mathcal{D} = 2$, that is at $h/J = 2J/U$. 

If instead $U \ll h$, it becomes inappropriate to consider $\mathcal{S}$ and $\mathcal{D}$ as almost disjointed subspaces, and the above transformation will be invalid. In this situation, the dominant effect will be coming from the strength of the quasiperiodicity $h$ and the interaction strength $U$ will become negligible, which implies that the system will be localised. Hence, after the transition that is due to interaction happens at $h/J = 2J/U$ for the subspace $\mathcal{D}$, the system remains localised as $h$ increases, where the dominant effect changes from interaction to quasiperiodicity. 

Note that for the subspace $\mathcal{S}$ the effective Hamiltonian to the lowest order equals to the original Hamiltonian $H_\text{sys}$ projected into $\mathcal{S}$. This is can be seen from \cref{eq:s1}, where the only relevant term that sits in $\mathcal{S}$ is $\lambda\hat{V}$. 

\bibliographystyle{rev10}
\bibliography{MBL}

\end{document}